\documentclass[a4paper,12pt]{article}

\usepackage{amsmath,amssymb,amsfonts}
\usepackage{physics}
\allowdisplaybreaks
\usepackage{graphicx}
\usepackage[dvipsnames]{xcolor}
\usepackage{caption}
\usepackage{subcaption}
\usepackage[bottom]{footmisc}	
\usepackage{soul}
\makeatletter
\@addtoreset{equation}{section}
\renewcommand{\theequation}{\thesection.\@arabic\c@equation}
\makeatother
\usepackage{hyperref} 
    \hypersetup{colorlinks=true, linktocpage, linkcolor={blue}, citecolor={blue}, urlcolor={blue}} 
\usepackage{cite}
\usepackage{caption}

\def \be {\begin{equation}}
\def \ee {\end{equation}}
\def \bea {\begin{align}}
\def \eea {\end{align}}
\def \nn {\nonumber}

\def \rr {\raise.35ex\hbox{\small $\prime$}\kern-.17em{\mbox{\large $\imath$}}}
\def \del {\partial}
\def \dels {\partial\kern-.5em / \kern.5em}
\def \As {{A\kern-.5em / \kern.5em}}
\def \Ds {D\kern-.7em / \kern.5em}

\def \a {\alpha}

\def \b {\beta}

\def \g {\gamma}

\def \d {\delta}

\def \s {\sigma}
\def \r {\rho}

\def \t {\tau}

\def \R {{\mathbb{R}}}

\newcommand{\cM}{{\mathcal M}}
\newcommand{\cB}{{\mathcal B}}
\newcommand{\cC}{{\mathcal C}}
\newcommand{\cH}{{\mathcal H}}
\newcommand{\cT}{{\mathcal T}}
\newcommand{\cL}{{\mathcal L}}
\newcommand{\cD}{{\mathcal D}}
\newcommand{\cR}{{\mathcal R}}
\newcommand{\cA}{{\mathcal A}}
\newcommand{\cF}{{\mathcal F}}
\newcommand{\cK}{{\mathcal K}}
\newcommand{\cG}{{\mathcal G}}

\def\TT{{\bf T}}

\def\R{{\mathbb R}} \def\C{{\mathbb C}} \def\N{{\mathbb N}}

\newcommand{\hs}{\mathfrak{hs}}

\def\und{\underline}

\def\End{\rm End}

\newcommand{\hide}[1]{}

\pdfoutput=1

\begin{document}



\begin{titlepage}
\vspace*{-10mm}   
\baselineskip 10pt   
\begin{flushright}   
\begin{tabular}{r} 
\end{tabular}   
\end{flushright}   
\baselineskip 24pt   
\vglue 10mm

\begin{center}

\noindent
\textbf{\LARGE
Modified gravity at large scales \\[1ex] 
on quantum spacetime 
in the IKKT model
    }
\vskip10mm
\baselineskip 20pt

\renewcommand{\thefootnote}{\fnsymbol{footnote}}

{\large
Harold C. Steinacker
}

\renewcommand{\thefootnote}{\arabic{footnote}}

\vskip5mm

{\it
Faculty of Physics, University of Vienna\\
Boltzmanngasse 5, 
A-1090 Vienna, Austria 
}

\vskip 20mm
\begin{abstract}
\normalsize

The gravitational dynamics of 3+1 dimensional covariant quantum spacetime in the IKKT or IIB matrix model is studied, combining the Yang-Mills-type matrix action with the induced Einstein-Hilbert action. This combined action leads to interesting modifications of the gravitational dynamics at long distances, governed by modified Einstein equations including an extra geometrical tensor interpreted as ''mirage matter''. In particular we find extra non—Ricci flat geometric modes with a non-standard dispersion relation, with features reminiscent of dark matter and/or dark energy.

\end{abstract}
\end{center}

\end{titlepage}

\pagestyle{plain}

\baselineskip 18pt

\setcounter{page}{1}
\setcounter{footnote}{0}
\setcounter{section}{0}

\tableofcontents



\section{Introduction}
\label{sec:Introduction}



The IKKT or IIB matrix model \cite{IKKT}
can be viewed as a non-perturbative formulation of type IIB string theory, where spacetime is thought
to emerge as a configuration of 9+1 matrices  $\TT_{\dot a}$.
The model is defined by the action 
\be
\label{IIB-action}
S_{\rm IIB} = \frac{1}{g^2}\mbox{Tr}\left([\TT_{\dot a}, \TT_{\dot b}] [\TT^{\dot a}, \TT^{\dot b}]
+  \bar{\psi}\Gamma^{\dot a}[\TT_{\dot a}, \psi]\right) \ 
\ee
where $\TT_{\dot a},\ \dot a = 0,...,9$ are Hermitian matrices, and $\psi$ are matrix-valued Majorana-Weyl spinors of  $SO(9,1)$. 
Assuming a non-trivial vacuum $\langle\TT_{\dot a}\rangle = \bar \TT_{\dot a}$  given by some  matrix background describing spacetime, the model
defines a gauge theory for the fluctuations $\TT_{\dot a} \to \bar\TT_{\dot a} + \cA_{\dot a}$. 
The dimension of that emergent spacetime is not given a priori, but there is mounting evidence from non-perturbative studies \cite{Anagnostopoulos:2022dak,Chou:2025moy,Chou:2024sgk,Asano:2024def} that 3+1 dimensions are dynamically preferred.
However, the specific structure of these matrices  is not evident. Note also that the coupling $g$ can be absorbed in the matrices or set to $g=1$, 
but we will keep to keep track of dimensions: matrices $\TT_{\dot a}$ will have dimension mass, and $g$
will have dimension $m^2$.

In \cite{Ho:2025htr} commutative matrix configurations were considered, given by
\be
\TT_{\dot \a} = P_{\dot \a}
\ee
where $P_{\dot \a}, \ \dot \a = 0,...,3$ are momentum generators acting on $\cH = L^2(\R^4)$.
The matrix model action can then be interpreted
as an action for bi-local fields. It was argued that this leads to general relativity (GR), assuming that a local theory emerges in the IR. Similar backgrounds with non-trivial geometry were considered in \cite{Hanada:2005vr}.
However, such backgrounds do not lead to a local action without further regularization.

In the present paper, we consider non-commutative matrices $[\TT_{\dot a}, \TT_{\dot b}] \neq 0$ as background. The underlying Hilbert space $\cH \cong L^2(\R^3)$ is much smaller  \cite{Ho:2025htr}, leading to a discrete set of normalizable higher-spin ($\hs$)-valued  modes rather than a continuous set of non-normalizable  $\hs$ modes. The resulting quantum spacetime carries only finitely many dof per volume and leads to a well-defined matrix model action for the fluctuation modes without regularization, even at one loop. 
Such noncommutative backgrounds  do not imply pathological zero modes, and it is 
plausible that they dominate the matrix (path) integral.

Specifically, we consider a class of quantum spacetimes $\cM^{1,3}$  where the global $SO(3,1)$ symmetry of the matrix model is equivalent to a gauge transformation. Such spaces are denoted as covariant quantum spaces and are expected to be preserved under quantum corrections. 
In contrast to basic quantum spaces such as $\R^{3,1}_\theta$, there is no $B$ tensor on spacetime.
The global spacetime geometry is of $k=-1$ FLRW type\footnote{There is a similar covariant quantum spacetime with curvature parameter $k=0$ \cite{Gass:2025tae}, which respects the global $E(3)$ symmetry of the model. } and has been studied in \cite{Sperling:2019xar,Battista:2023glw}. Nevertheless, many of our results will generalize to other quantum spacetimes.

The fluctuations on noncommutative backgrounds in the matrix model are governed by a noncommutative Yang-Mills type gauge theory. However,
the mechanism for gravity is not evident. Although the classical action defines a dynamics for noncomutative spacetime, it is different from the Einstein-Hilbert action \cite{Steinacker:2010rh}. 
The latter arises only after quantization, as confirmed in an explicit 1-loop computation in \cite{Steinacker:2023myp}. 
In other words, {\em gravity is a quantum effect} on quantum spacetime.
However, the classical Yang-Mills-type action does not fit into the framework of general relativity. 
One way to address this issue was proposed in \cite{Kumar:2023bxg} by introducing an "anharmonicity tensor", which 
captures the geometric sector of the Yang-Mills action
non-locally in terms of the frame.

In the present paper, we study the  gravitational theory obtained by combining the Einstein-Hilbert term with a geometrical Yang-Mills action. 
Our aim is to clarify the relation and distinction of the resulting gravity theory from GR. 
We find a modification of GR by a non-local sector, leading to 
new gravitational physics at very large distances.
Non-locality arises because
the fundamental matrix degrees of freedom act as {\em potentials} for the frame (or metric). 
We obtain tensorial equations describing the modified gravitational dynamics in terms of an "anharmonicity tensor", which appears as a non-local "mirage" of actual matter, but can acquire its own dynamics.


These modifications of GR in the IR arise from extra gravitational modes which are not Ricci-flat, and satisfy a non-relativistic dispersion relations. These modes are very weakly coupled to (or generated by) the local matter distribution, but they can lead to significant effects at large scales. In particular, static localized matter induces an extra deformation of the metric beyond GR, which may be interpreted in terms of a halo-like mirage (or dark) matter distribution with characteristic scale $m_{\rm cross}$. This is a dynamical IR scale where general relativity is  modified. We find that this scale is separated hierarchically from the Planck scale and the UV scale of spacetime, but a reliable assessment is beyond the present understanding of the model and the background.

These modifications of gravity goes beyond the more standard deformations such as $f(R)$ gravity, but there are some similarities with massive gravity.
While the extra modes appear to suffer from an IR instability at the linearized level, we argue that this instability will be cured in a more complete form of the effective action. This also involves the $\hs$ sector of the theory, which is an intrinsic feature of covariant quantum spacetime. 

A related finding is the lack of manifest local Lorentz invariance in the novel sector. The local, tensorial sector of the resulting theory is covariant and hence respects local Lorentz invariance, but the extra non-local sector of the theory is governed by a metric with reduced speed of propagation. Since
the frame is a derived object which arises from a potential and is divergence-free, it does not admit local Lorentz transformations. Nevertheless, the model admits a gauge symmetry interpreted as higher-spin extension of volume-preserving diffeos, which ensures a ghost-free fluctuation spectrum \cite{Steinacker:2019awe}.

Finally, the present approach should not be confused with a holographic point of view, which aims to identify some effective $9+1$ dimensional target space geometry \cite{Komatsu:2024bop,Hartnoll:2024csr,Ciceri:2025wpb}. Here we focus on the {\em intrinsic} $3+1$ dimensional geometry and physics of some background or "brane", which is assumed to be sufficiently stable and perhaps dominant.

\section{Minimal covariant quantum spacetime}

We first describe the algebraic structure of the covariant quantum spacetime following \cite{Sperling:2019xar,Manta:2025inq}, which is considered as a background for the matrix model.

\subsection{Spacetime algebra and realization}

Covariant quantum spacetime is based on a unitary irreducible "doubleton" representation $\cH_n$ of $SO(4,2)$, for $n\in\N$ \cite{Govil:2013uta}. We will restrict ourselves to the minimal case $n=0$. The full algebra of operators is generated by 4+4 generators $X^\mu, T^\nu$ with commutation relations
\begin{align}
[T^\mu,X^\nu] &= \frac ir X_4 \,\eta^{\mu\nu}\nn\\
 [X^\mu,X^\nu] &= -ir X_4^{-1}(T^\mu X^\nu - T^\nu X^\mu)\ ,  \nn\\
 [T^\mu,T^\nu] &= i X_4^{-1}(T^\mu X^\nu - T^\nu X^\mu) \ 
  \label{XT-CR}
\end{align}
(for $n=0$), where $X_4^2 = r^2 -  X_\mu X^\mu$. These generators
arise from the $SO(4,2)$ Lie algebra as $X^a = r M^{a 5}$ and  $T^\mu = r^{-1} M^{\mu 4}$, and hence are covariant under $SO(3,1)$.
The first bracket relation suggests to interpret the $T^\mu$ as momentum generators acting on functions of $X^\nu$, while the latter are interpreted as spacetime generators. The doubleton representations $\cH_n$ imply two additional algebraic constrains,
which for $n=0$ take the form
\begin{align}
   r^2 T_\mu T^\mu &= - r^{-2} X_\mu X^\mu \ , \qquad
    X_\mu T^\mu + T_\mu X^\mu = 0 \ .
    \label{TT-XX-relation}
\end{align}
These allow to express $T^0$ and $X^0$ as functions of the 3+3 independent generators $X_i$ and $T_i$.
The operator algebra $\End(\cH_0)$ can be interpreted as quantized algebra of functions on  a 6-dimensional symplectic space\footnote{The underlying sympletic space $\cB$ can be recognized as twistor space.}  $\cB$, viewed as $S^2$ bundle over cosmological spacetime $\cM^{3,1}$:
\begin{align}
    \End(\cH_0) \cong  \cC(\underbrace{\cM^{3,1} \times S^2}_{\cB})
    \cong  \cC(\cM^{3,1}) \otimes \hs
\end{align}
Due to the constraints, the trace over $\End(\cH)$ reduces locally to an integral over $\cM^{3,1} \times S^2$, with only {\bf finitely many dof per volume }\cite{Manta:2025inq}. This is essential for the matrix model to be well-defined; without the constraints, the trace would be UV-divergent.
We will provide an elementary description of these relations in the semi-classical approximation.

The generic case with $n\geq 1$ reduces to the above relations in the  late-time regime $X^0 \gg r$. We will therefore only consider the case $n=0$ in the following.

We consider the matrix model $S_{\rm IIB}[{\bf T}^{\dot a}]$ for perturbations of the background defined by
\begin{align}
\label{matrix-background-full}
    {\bf T}^{\dot a} = \left\{ \begin{array}{cl}
        T^{\dot \mu} &, \  \dot \mu = 0,1,2,3  \\
        0 &, \ \ \dot a = 4,...,9 
    \end{array}   \right.
\end{align}
or its generalization \eqref{deformed-background}. This background is covariant under $SO(3,1)$ symmetry of the matrix model, 
\begin{align}
\label{covariance}
    U^{-1} T^{\dot \mu} U = \Lambda^{\dot \mu}_{\ \dot \nu} T^{\dot \nu} \ 
\end{align}
i.e. the global symmetry is equivalent to a gauge symmetry. Such covariant backgrounds are clearly preferred in the quantized matrix model, i.e. upon integrating over all matrices.
To obtain gravity, the transversal matrices ${\bf T}^{\dot a},\ \dot a = 4,...,9$ must in fact be non-trivial and describe some fuzzy extra dimensions $\cK$, which can be stabilized by $R$ charge \cite{Manta:2025tcl}. We will mostly ignore this for simplicity here.


\subsection{Semi-classical description}

 The eigenvalues of the time-like matrix $X^0$ on $\cH_0$ are given by the positive integers,
 \begin{align}
 \label{X0-spec}
 {\rm spec}(X^0)  \ = \ r\{1,2,3,... \} \ .
 \end{align}
At late times $X_0 \gg r$, this quantum spacetime can be described approximately by a Poisson algebra
on a symplectic space $\cB \cong \cM^{3,1} \times S^2$,
with $4 + 4$ generators $x^\mu \sim X^\mu$ and $t_\nu \sim T_\nu$ satisfying the  bracket relations
\begin{align}
\{t^\mu,x^\nu\} &=  \frac{x_4}r \,\eta^{\mu\nu}\nn\\
 \{x^\mu,x^\nu\} &= - \frac{r^3}{x_4}(t^\mu x^\nu - t^\nu x^\mu)\ =: \theta^{\mu\nu} \ ,  \nn\\
 \{t^\mu,t^\nu\} &= \frac{1}{r x_4} (t^\mu x^\nu - t^\nu x^\mu) \
  \label{XT-CR-semi}
\end{align}
 and the constraints
\begin{align}
   r^2 t_\mu t^\mu &= - r^{-2} x_\mu x^\mu \ , \qquad
  t_\mu x^\mu = 0 \quad \mbox{i.e.} \ \ t^0 = \frac{1}{x^0} t_i x^i \ .
    \label{TT-XX-relation-2}
\end{align}
It is easy to verify the Jacobi identity.
We also define
\begin{align}
  x_4^2 = r^2 -  x_\mu x^\mu \ \approx  -  x_\mu x^\mu \ .
\end{align}
From now on we will work in the semi-classical regime where commutators can be approximated by Poisson brackets, and denote this Poisson algebra by $\cC$.

\begin{figure*}[t!]
    \centering
    \includegraphics[scale=0.52]{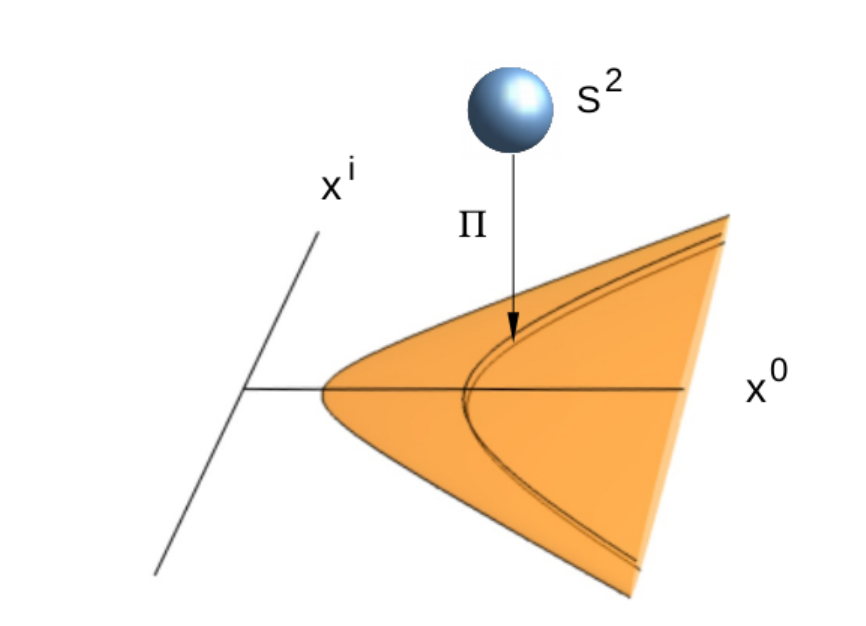}
    \caption{Sketch of the bundle space $\cM^{3,1}\times S^2$. The black lines indicate space-like $H^3$.}
    \label{fig:bundle}
\end{figure*}

To understand the geometrical meaning of this Poisson algebra, consider
some reference point $x = \xi = (\xi^0,0,0,0)$ on $\cM^{3,1}$ with $\xi^0 \gg 1$. Then the constraints reduce to $t_\mu t^\mu = \xi_0^2$ and
$t^0 = 0$, so that $t^\mu$ is a space-like vector
of length $\xi_0$.
Therefore $\cC$ describes the algebra of functions on an $S^2$ bundle over spacetime, with $S^2$ generated by the space-like $t_i$ with radius $\xi_0^2$.
Spacetime itself can be recognized as a (double cover of) a $k=-1$ FLRW spacetime with Big Bounce at $x_4 =0$.
A schematic picture of this space is shown in figure \ref{fig:bundle}.
The algebra of functions thus
decomposes into a tower of spin $s$ valued functions
\begin{align}
    \cC = \C[[x^\mu,t_\nu]]/_\sim \
    = & \ \  \cC^0 \ \oplus \ \cC^1 \ \oplus \ \cC^2
 \oplus ...  \nn\\
 & \ \ \cC^s \ni \phi^{(s)} = \phi_{\und{\mu}}(x) t^{\und{\mu}}
\end{align}
(here $\und{\mu}$ is a multi-index)
in terms of irreducible polynomials in $t^\mu$ of degree $s$.
Due to the constraint $t_\mu x^\mu = 0$ the $t^\mu$ are space-like\footnote{this can be seen near the reference point $\xi$.}, so that the tensor fields $\phi_{\und{\mu}}(x)$ can be chosen to be space-like. This will be denoted as {\bf unitary gauge}, where the absence of ghosts is manifest. 
The projection  to the classical (spin 0) sector $\cC^0$ is achieved  by 
\begin{align}
\label{proejction-tt}
    [t_\mu t_\sigma]_0 = \frac{1}{3r^2} \kappa_{\mu\sigma}, \qquad
    \kappa_{\mu\sigma} = \frac 1{r^2}\big(x_4^2 \eta_{\mu\sigma} + x_\mu x_\sigma\big) \ .
\end{align}
Note that $\kappa_{\mu\sigma}$ is a space-like tensor, since $x^\mu [t_\mu t_\sigma]_0 = 0$.

\paragraph{Effective geometry.}

In the semi-classical regime,
the kinetic term in the action for {\em all} fluctuations of the matrices around some (generic) background $\TT^{\dot\a}$ in the matrix model is governed by an effective metric $G_{\mu\nu}$, which arises from a  frame
\begin{align}
    e^{\dot\a \mu} =  \{\TT^{\dot\a},x^\mu\} = e^{\dot\a \mu}(x) \ .
\end{align}
Locally, all $\hs$ components can be eliminated\footnote{This requirement determines the local algebra of functions on spacetime generated by $x^\mu$, and it can always be achieved locally as shown in appendix \ref{sec:LNC}. This holds identically for the covariant background.}.
This defines the effective metric and the dilaton
\cite{Steinacker:2010rh,Sperling:2019xar,Steinacker:2024unq}:
\begin{align}
G_{\mu\nu} &:= \rho^2 \gamma_{\mu\nu}, 
\qquad \gamma^{\mu\nu} = \eta_{\dot\a\dot\b} e^{\dot\a \nu}e^{\dot\b \nu}  \label{eff-metric-def}\\
 \rho^2 &= \rho_M \sqrt{|\gamma^{\mu\nu}|}
 = \rho_M \det(e^{\dot\a\mu})
  = \frac{\sqrt{|G|}}{\rho_M}
  \label{dilaton-def}
\end{align}
where $\rho_M$ is the volume form on $\cM^{3,1}$ which arises by integrating the symplectic volume form over $S^2$.
Here and in the following, dotted greek indices indicate frame indices, which transform under the global $SO(3,1)$ symmetry of the matrix model.
The effective metric is encoded in the (semi-classical) matrix Laplacian
\begin{align}
    \Box = -\{\TT^{\dot\a},\{\TT_{\dot\a},.\}\}
    &= \rho^2 \Box_G 
     = \frac{\rho^2}{\sqrt{|G|}}\del_\mu\big(\sqrt{|G|} G^{\mu\nu}\del_\nu . \big) 
    \approx -\gamma^{\mu\nu}\del_\mu\del_\nu \ .
\end{align}
In particular, the covariant background $\TT^{\dot\a} \sim t^{\dot\a}$ 
defines the frame
\begin{align}
    e^{\dot\a \nu} = \{t^{\dot\a},x^\nu\}
    = \sinh(\t)\, \eta^{\dot\a \nu}
\end{align}
in Cartesian coordinates, where 
$\tau$ is a convenient time parameter $\tau$ on $\cM^{3,1}$ defined by
\begin{align}
    x_4 = r\sinh(\tau) \ .
\end{align}
Then the effective metric is obtained as
 \cite{Sperling:2019xar}
\begin{align}
G_{\mu\nu} &:= \ r^{-4}\sinh(\t)\,\eta_{\mu\nu}, 
\qquad \gamma^{\mu\nu}  \approx \sinh^2(\t)\,\eta^{\mu\nu} \ .
\end{align}
 This describes a $k=-1$ FLRW geometry with
Big Bounce at $\tau=0$ and
spacelike $SO(3,1)$ isometry acting on the background via \eqref{covariance}.  The global one-form $d\tau$ dual to the time-like vector field
$\cT = \partial_\tau$ is compatible with $SO(3,1)$.
Note that $r$ should be considered as a number here.
The symplectic volume form 
governs the quantization of 4-volume, and
is related to the trace via \cite{Manta:2025inq}
\begin{align}
\label{trace-semiclass}
    \Tr \sim \int \Omega =  \int d^4 x \rho_M \ \sim \int d^4x \sqrt{G}\rho^{-2}, 
    \qquad
 \rho_M &\sim \frac 1{2\pi^2 r^3 x^4}  \sim \frac {e^{-\t}}{\pi^2 r^{4}}  \ .
\end{align}
A similar background leading to a $k=0$ FLRW geometry is discussed in \cite{Gass:2025tae}.

To describe the 1-loop action, it is useful to consider the Weitzenb\"ock torsion associated to the frame $e^{\dot\a\mu}$, which is defined as \cite{Steinacker:2020xph}
\begin{align}
\label{Weitzenbock-torsion-def}
      T^{\dot\a\dot\b\mu} &= e^{\dot\b}[e^{\dot\a\mu}] - e^{\dot\a}[e^{\dot\b\mu}] 
       = \{\cF^{\dot\a\dot\b},x^\mu\}\ , \qquad \cF_{\dot\a\dot \b} = -\{\TT_{\dot\a},\TT_{\dot\b}\} \ . 
\end{align}
Here $\cF_{\dot\a\dot \b}$ is the Yang-Mills field strength, which defines  the classical matrix model
\cite{Aoki:1999vr}. The totally antisymmetric part of this torsion defines an "axionic" vector field \cite{Fredenhagen:2021bnw,Kumar:2023bxg}
\begin{align}
\label{axion-def}
  \tilde T_{\sigma}\ :=\ -\frac{1}{3!}\sqrt{|G|}\,\varepsilon_{\nu\mu\rho\sigma}G^{\mu\mu'}G^{\rho\rho'}\,T^{\textrm{(AS)}\nu}_{\qquad\mu'\rho'}\ .
\end{align}

\subsection{Local description of covariant quantum spacetime}
 
This section provides a simplified description of the above quantum spacetime, zooming into the locally flat regime.
Consider again the reference point $\xi = (\xi^0,0,0,0)$ with $r^{-1}\xi^0 =  \cosh(\tau) \gg 1$.
Near this reference point $\frac{x^i}{\xi^0} \ll 1$ and $\frac{x^4}{x^0} \approx 1$, and either $x^4$ or $x^0$ can be considered as time parameter. We
define local coordinates $y^\mu$ near $\xi$ via
\begin{align}
    x^\mu = \xi^\mu + y^\mu \ .
\end{align}
Then the above relations take the form
\begin{subequations}
\label{y-t-brackets}
\begin{align}
 \{t^\mu,y^\nu\} &\approx  \frac{\xi_4}{r}\eta^{\mu \nu} ,\qquad  \nn\\
\{y^0,y^i\} 
&\approx r^3 t^i  , \qquad \ \ \ 
 \{y^i,y^j\} =  \frac {r^3}{\xi_4}(y^i t^j - y^j t^i)
 = O(r y^i)\  \nn\\
  \{t^0,t^i\} 
  &\approx  -  r^{-1} t^i , \qquad \{t^i,t^j\}  = -  \frac 1{r\xi_4}(y^i t^j - y^j t^i) 
  = O(r^{-3} y^i) \   
 \end{align}
\end{subequations}
using $\xi_0 = \xi_4$.
Hence the $\{t_\mu,.\} \sim \del_\mu$ play the role of derivatives acting on functions of $y^\mu$.
Near the reference point, the constraints reduce to
\begin{align}
    t^0 &= \frac{\vec y \cdot\vec t}{\xi^0} \approx 0  \nn\\
  r^4\vec t^2 &= (x^0)^2 - \vec y^2 +(t_0)^2 \approx \xi_0^2
\end{align}
assuming $\frac{|\vec y|}{\xi^0} \ll 1$. This means that $\vec t$ describes a space-like sphere with radius $r^{-2}\xi^0$.
It is convenient to use normalized $\hs$ generators
\begin{align}
 u^\mu = \frac{r^2}{x_4} t^\mu \ , \qquad u_\mu u^\mu = 1 \ .
\end{align}
Since $u^0 \approx 0$ (as a function on the bundle $\cB$), they reduce to
\begin{align}
u^i &= \frac{r^2}{\xi_0}t^i , \qquad\qquad  u^i u_i = 1 \ .
\label{u-tilde-p}
\end{align}
The most general functions can now be expanded in a discrete sum of higher-spin modes
\begin{align}
\phi = \sum_s \phi_{sm} \hat Y^{sm}(u)
\end{align}
where $\hat Y^{sm}(u)$ are irreducible polynomials in $u^i$ i.e. spherical harmonics on the internal $S^2$.
The projection \eqref{proejction-tt} takes the form
\begin{align}
\label{u-average}
 [u_i u_j]_0 &= \frac 1{3}  \delta_{ij} , \qquad [u_0 u_j]_0 = 0 , \qquad [u_0 u_0] = 0
\end{align}
near $\xi$.
Then the bracket relations can be written locally as\footnote{These relations coincide with those for $k=0$ covariant quantum spacetime \cite{Gass:2025tae} in the local regime.}
\begin{align}
\boxed{
\begin{aligned}
 \{u^\mu,y^\nu\} &\approx r(\eta^{\mu \nu} - u^\mu u^\nu) ,  \\
\{y^0,y^i\} &= L_{NC}^2 u^i \\
 \{y^i,y^j\} &= r(y^i u^j - y^j u^i)
\ll  L_{NC}^2 \\
   \{u^0,u^i\} &\approx 0 \\
 \{u^i,u^j\} &\approx 0 \ .
 \end{aligned}
 } 
 \label{y-u-brackets}
\end{align}
Here
\begin{align}
 L_{\rm NC} \ = \sqrt{r\xi_0} 
\end{align}
 is the local scale of noncommutativity in the $y^\mu$ coordinates, which is encoded in
 \begin{align}
 \label{theta-munu-explicit}
\theta^{\mu\nu} :=  \{y^\mu,y^\nu\} &\approx r(\xi^\mu u^\nu - \xi^\nu u^\mu)
 = O(L_{\rm NC}^2) \ .
 \end{align}
We note that the space-like $y^i$ can be considered as commuting.
More generally we will encounter three different length scales
\begin{align}
\label{L-NC}
    L_{\rm UV} := r \ll L_{\rm NC} \ = \sqrt{r\xi_0} \ \ll \   L_{\rm IR} = \xi_0 \ 
\end{align}
(in Cartesian coordinates).
Here  $L_{\rm UV}$ is the fundamental quantization scale of time $y^0$ \eqref{X0-spec}, 
$L_{\rm IR}$ a cosmic scale, and $L_{\rm NC}$ is the geometric mean of them. 
$L_{\rm IR}$ also sets the
{\bf locally flat regime} of the background in the following sense:
For functions $\phi(y)$ with wavelengths much smaller than $L_{\rm IR}$, the brackets with the functions  dominate the brackets between single generators, so that e.g.
\begin{align}
\label{locally-flat-estimate}
    \{\phi(y) u_\mu,y^\nu\} &=  \{\phi(y) ,y^\nu\} u_\mu + \phi(y)\{u_\mu,y^\nu\}  \nn\\
    &\approx 
      \{\phi(y) ,y^\nu\}u_\mu = O\big(L_{\rm NC}^2 \del \phi\big)  \nn\\
\{\phi(y) u_\mu,u^\nu\} 
    &\approx  \{\phi(y) ,u^\nu\}u_\mu
    =O(L_{\rm UV}\del \phi) 
\end{align}
for $\phi(y) u_\nu \in \cC^1$. Hence the $u^\mu$ can be treated as commuting in this regime.

The algebraic structure \eqref{y-t-brackets} is expected to hold in local normal coordinates on generic
deformed backgrounds corresponding to non-trivial gravitational fields, 
within a local coordinate patch of size given by the gravitational curvature scale \cite{Steinacker:2024unq}.

\paragraph{Divergence-free frame and metric.}

Now consider a generic matrix  configuration 
\begin{align}
\label{frame-def}
    {\bf T}^{\dot\a} = t^{\dot\a \nu}(y) u_\nu \quad \in \cC^1 
\end{align}
considered as a deformation of the background ${\bf T}^{\dot\a} = t^{\dot\a}$ or \eqref{deformed-background}.
It acts on functions as 
\begin{align}
\label{frame-def-2}
    \{{\bf T}^{\dot\a},\phi(y)\} &= e^{\dot\a\mu}\del_\mu\phi \ , \qquad e^{\dot\a\mu} := \{{\bf T}^{\dot\a},y^\mu\}
\end{align}
and thereby defines a frame $e^{\dot\a\mu}$ in the matrix model, which is  automatically divergence-free
\begin{align}
\del_\mu (\rho_M e^{\dot\a\mu}) = 0
\end{align}
due to the Jacobi identity \cite{Fredenhagen:2021bnw}.
Conversely, any divergence-free classical vector field or frame $e^{\dot\a\mu}_0$ can be written as  $e^{\dot\a\mu}_0 = [\{\TT^{\dot\a},y^\mu\}]_0$ for some (uniquely determined) generators $\TT^{\dot\a} =  t^{\dot\a \mu}(y) u_\mu   \in \cC^1$ (see Appendix \ref{sec:hs-Lorentz} for a simple argument).
Moreover, all $\hs$ components of $e^{\dot\a\mu}$ can always be eliminated locally using suitable coordinates, as explained in Appendix \ref{sec:LNC}.

We observe that the most general divergence-free frame encodes 12 (off-shell) degrees of freedom, just like the matrix background \eqref{frame-def}. 
Any metric $\gamma^{\mu\nu}$ can be realized (at least in some local patch) in terms of a divergence-free frame, accounting for 10 dof; note that 
the dilaton is constrained through to \eqref{dilaton-def}.
The remaining 2 dof are naturally attributed to the axionic vector field \eqref{axion-def}, which can be written on-shell as $\tilde T_\mu = \rho^{-2}\del_\mu\tilde\rho$
\cite{Fredenhagen:2021bnw}.

\subsection{Generalized cosmological background}

To accommodate the time evolution imposed by the matrix model, 
 we consider backgrounds with a generic scale function $\a(\t)$:
\begin{align}
\label{deformed-background}
    {\bf T}^{\dot\a} = \a(\tau) t^{\dot\a} \ .
\end{align}
This is the most general background which respects the $SO(3,1)$ covariance \eqref{covariance} (at any given $\tau$) \cite{Battista:2023glw}, therefore it is expected to be preserved by quantum corrections.
For the local physics, this amounts to replacing $r \to \frac{r}{\a}$ for the background, however we will continue to use the un-modified local Cartesian coordinates\footnote{This ignores some subtleties for the $\hs$ components \cite{Manta:2025tcl}, which do not entail essential changes.}  $y^\mu$ around the reference point $\xi = r(\cosh\tau,0,0,0)$. 

\subsection{Metric and scales}

The above background defines a frame as
\begin{align}
\label{frame-e-gen}
    e^{\dot\a\mu} &= 
    \{{\bf T}^{\dot\a},y^\mu\}  
    \approx \a \sinh(\t) \eta^{\dot\a\mu} \nn\\
    E^{\dot\a\mu} &= \rho^{-1} e^{\dot\a\mu} 
\end{align}
in Cartesian coordinates (treating $\a(\tau)$ as "almost"- constant\footnote{This is justified locally, see \cite{Manta:2025tcl} for a careful justification in local normal coordinates.}), which is automatically divergence free.
The dilaton 
measures the Riemannian volume per  quantum cell, given by 
\begin{align}
\label{rho-eplicit-cartes}
   \rho^2 &= \frac{\sqrt{G}}{\rho_M} 
   = \rho_M \det (e^{\dot\a\mu})
    \sim  e^{3\t} \frac{\a^4}{r^4}  
\end{align}
where $\rho_M \sim r^{-4} e^{-\tau}$ is the symplectic density \eqref{trace-semiclass} in local Cartesian coords $y^\mu$ near $\xi$. 
It also sets the Yang-Mills coupling constant as $g^2_{\rm YM} = \frac{g^2}{\rho^2}$, which is dimensionless.
Then the effective metric near $\xi$ is obtained as
\begin{align}
G^{\mu\nu} &= 
\rho^{-2} \eta_{\dot\a \dot\b}e^{\dot\a\mu} e^{\dot\b\nu} 
 = \frac{r^4}{\a^2\sinh(\tau)} \eta^{\mu\nu},   \
\label{eff-metric-App} 
\end{align}
leading to a cosmic FLRW scale function  (see appendix \ref{sec:cosm-scale})
\begin{align}
     a(t) \sim  \frac{\alpha}{r} e^{3\tau/2}
     \sim  t
    \label{at-modified-NC-1}
\end{align}
at late times, provided $\a\approx const$. This describes an expanding universe  with FLRW time parameter $t$, with Hubble scale \eqref{Hubble-scale-app}
\begin{align}
    H(t) &=  \big(\varepsilon + \frac 32\big) \frac 1{a(t)} \sim  \frac 1t \ .
\end{align}
This is the curvature scale of the universe at time $\tau$ measured by $\xi^\mu \xi^\nu G_{\mu\nu} \sim \frac{\a^2}{r^2}e^{3\tau}$.
Similarly, the uncertainty scale  \eqref{L-NC}
\begin{align}
 L_{\rm NC} =  r e^{\t/2}
\end{align}
in Cartesian coordinates $y^\mu$ 
corresponds to an effective length scale 
\begin{align}
\label{LNC-a-estimate}
  L_{\rm NC,G}^2 =   \Delta y^\mu\Delta y^\nu   G_{\mu\nu} = \frac{\a^{2}}{r^2} e^{2\t} \ 
\end{align}
which depends only on $\a/r$.
We observe an interesting {\em dimensional transmutation}: 
$r$ turns out to have dimension ${\rm length}^{-1}$,
 since the metric tensor \eqref{eff-metric-App} contains itself $r$ as a parameter, and it will happen consistently in the following. Of course $r$ is really a number, and dimensionful quantities only arises in the effective physics of fluctuations, as identified above.

There is a much higher UV scale which is presumably more meaningful:  $L_{\rm NC}$
arises from $\hs$ components in $\{x^\mu,x^\nu\} \sim \theta^{\mu\nu}$ \eqref{XT-CR-semi}, but these are suppressed upon evaluating in an $SO(3)$-invariant state, where the uncertainty of $x^i, i=1,...,4$ is reduced to 
\begin{align}
    L_{UV}^2 := \langle\psi_0|\theta^{ij}|\psi_0\rangle = O(r^2e^{\frac 12\tau}) \ll L_{NC}^2, \qquad i,j=1,...,4
\end{align}
 (i.e. averaging over the $S^2$ fiber), consistent with the 4-volume quantization ${\rm vol}_\Omega B^4_R  \propto N^3 \propto e^{3\tau}$ via $\rho_M$ for $R=Nr \propto r e^\tau$ \eqref{trace-semiclass} \cite{Manta:2025inq}. This is expected the UV scale where QFT breaks down, which should be elaborated elsewhere.
In terms of the effective metric, this scale is 
\begin{align}
\label{UV-scale-new}
      L_{\rm UV,G}^2 = \frac{\a^{2}}{r^2} e^{\frac 32\t} \ = e^{-\frac 12\tau} L_{\rm NC,G}^2 \ .
\end{align}
Finally the cosmic scale defined by the inverse Hubble rate is
\begin{align}
\label{Hubble-scale}
   L_{\rm cosm} = \frac 1{H(t) } & \sim  \  \frac{\alpha}r e^{3\tau/2}  \ \gg L_{\rm NC,G}\sim  \ \frac{\a}r e^{\t}  \ \gg \ 
   L_{\rm UV,G}\sim  \ \frac{\a}r e^{\frac 34\t}
\end{align}
consistent with the above dimensional assignment. 
These 3 scales are separated by a large factor $e^{\tau/2}$ and $e^{\tau/4}$
at late times, as it must be.
Note that $L_{UV,G}$ characterizes the fundamental UV scale of spacetime, while $L_{\rm NC,G}$ characterizes the scale where the $\hs$ structure in the time-space commutators becomes relevant.

\section{Modified Einstein equations and gravity}
\label{sec:mod-einstein}

Now we can derive the modified Einstein equations arising from the matrix model at one loop. 
The gravity sector of the one-loop contribution to the IKKT model is \cite{Steinacker:2023myp}
\begin{align}
\label{S-grav-1loop}
    S_{\rm grav} = - \frac{1}{2} \int
 \frac{ d^4x \sqrt{G}}{16\pi G_N} \rho^{-2}\, T^{\dot\a\dot\b\mu} T_{\dot\a\dot\b}^{\ \ \mu} \gamma_{\mu\nu}
 = \int
 \frac{ d^4x \sqrt{G}}{16\pi G_N} 
 \big(\cR[G] + \frac{1}{2}\tilde T\cdot \tilde T - 2\rho^{-2}\del\rho\cdot\del\rho \big)\ 
\end{align}
where $G_N$ is the effective Newton constant\footnote{The scale of the Newton constant is set by the Kaluza-Klein scale of the fuzzy extra dimensions $\cK$ described by the extra 6 matrices in \eqref{matrix-background-full}. We refer to \cite{Steinacker:2023myp} for more details.}, and $\tilde T^\mu$ is the axionic vector field \eqref{axion-def}.
For simplicity we will drop the axionic and the dilaton contribution on the rhs, and vacuum energy will be included later.
Then the variation of the E-H term takes the  form \cite{Kumar:2023bxg}
\begin{align}
    \delta S_{\rm E-H}  = &\int\frac{\sqrt{G}}{8\pi G_N} \rho^{-2}\cR_{\mu\lambda}\, 
 e^{\mu}_{\dot\a} 
    \delta e^{\dot\a\lambda}\ 
\end{align}
in terms of the frame. This is expected to hold very generally, beyond the specific 1-loop calculation.
That term should be combined with the bare Yang-Mills-type action\footnote{This is the $3+1$ section of the matrix model action \eqref{IIB-action}. The coupling constant $g^2$ of the model is absorbed in the background.} 
\begin{align}
\label{S-YM}
    S_{\rm YM} =  -\frac 1{g^2}\int
 d^4y \sqrt{G}\rho^{-2} \cF^{\dot\a\dot\b} \cF_{\dot\a\dot\b} \ , \qquad
 \cF_{\dot\a\dot \b} = -\{\TT_{\dot\a},\TT_{\dot\b}\} \ 
\end{align}
with variation 
\begin{align}
\label{delta-SYM-0}
  \delta S_{\rm YM} & = -\frac 4{g^2}\int\ d^4x\rho_M\, \d \TT_{\dot\alpha}\, \Box \TT^{\dot\a}  
  = \frac 4{g^2}\int d^4x \rho_M \d \TT_{\dot\alpha}\,\{\TT^{\dot\b},\cF_{\dot\a\dot\b}\} \ .
 \end{align} 
 To combine these two contributions, 
 we rewrite the latter in terms of 
an "anharmonicity tensor" $C_{\dot\a\mu}$ defined by \cite{Kumar:2023bxg}
\begin{align}
\label{anharm-tensor-def}
    \{C_{\dot\a\mu},x^\mu\} = - \frac 1{g^2}e^{\dot\b}[\cF_{\dot\b\dot \a}] \ .
\end{align}
This will be solved in \eqref{C-explicit-formula-1}; 
the solution is not unique, but admits the following  ambiguity or "gauge invariance" 
\begin{align}
 C_{\dot\a\mu} \to C_{\dot\a\mu} + \partial_\mu C_{\dot\a}   \ .
 \label{C-ambiguity}
\end{align}
Then
 \begin{align}
  \delta S_{\rm YM} &= 4\int d^4x \rho_M \d \TT_{\dot\alpha}\,\{C_{\dot\a\mu},x^\mu\}  
  = - 4\int d^4x \rho_M \{\d \TT_{\dot\alpha},x^\mu\}  \, C_{\dot\a\mu} \nn\\
  &= - 4\int d^4x \sqrt{G} \rho^{-2}\, \d e_{\dot\alpha}^\mu  \, C_{\dot\a\mu} \nn\\
 &\equiv  - \int d^4x\sqrt{G}\rho^{-2}\,
 \big(T_{\mu\nu}[C] - \frac 12 G_{\mu\nu}T[C] + B_{\mu\nu}[C]\big)
 e_{{\dot\a}}^{\ \mu} \d e^{{\dot\a}\nu}
\end{align}
where
\begin{align}
\label{TC-B-def}
T_{\mu\nu}[C] - \frac 12 G_{\mu\nu} T[C]
+ B_{\mu\nu}[C] :=  4 C_{\dot\a\nu} e^{\dot\alpha}_\mu 
\end{align}
(the antisymmetric part $B_{\mu\nu}$  will be dropped for simplicity in the following).
Comparing with the variation of the matter action in terms of the frame
\begin{align}
\label{varMatter}
    \delta S_{\textrm{matter}} & = - \frac{1}{2}\int \sqrt{|G|}\,T_{\mu\nu}\delta G^{\mu\nu} 
    = - \int \sqrt{|G|}\,\rho^{-2} \big(T_{\mu\lambda} - \frac 12 G_{\mu\lambda}T \big) e^\mu_{\dot\a} \delta e^{\dot\a\lambda}\ ,
\end{align}
with  $T = G^{\mu\nu}T_{\mu\nu}$, we can interpret $T_{\mu\nu}[C]$ as an effective energy-momentum tensor, interpreted as apparent "mirage" (dark) matter. 
Hence the modified Einstein equations are\footnote{
The divergence constraint $\del_\mu e^{\dot\a\mu} = 0$ for the frame should be taken into account in terms of Lagrangian multipliers $C^{\dot\a}$, which leads precisely to the ambiguity \eqref{C-ambiguity} of $C_{\dot\a\nu}$. The 
present discussion is simplified dropping dilaton and axion contributions, see \cite{Kumar:2023bxg} for a more detailed computation.} 
\begin{align}
  \label{mod-Einstein-frame}
\frac 1{8\pi G_N} \cR_{\mu\nu} e^\mu_{\dot\a}
= \big(T_{\mu\nu} - \frac 12 G_{\mu\nu}T \big) e^\mu_{\dot\a}
 + 4C_{\dot\a\nu}
\end{align}
or
\begin{align}
  \label{mod-Einstein}
  \boxed{
\frac 1{8\pi G_N} \cG_{\mu\nu} 
= T_{\mu\nu} + T_{\mu\nu}[C] \ . \
 }
\end{align}
This equation will be supplemented by a non-local relation between $ T_{\mu\nu}[C]$ and matter \eqref{T-C-expression-metric-2}.
The resulting modification of 
gravity is discussed below.
Note that $C_{\dot\a\mu}$ also absorbs the $\hs$ valued components of $\cG_{\mu\nu}$, and
the ambiguity \eqref{C-ambiguity} is used 
to make $T^{\mu\nu}[C]$ divergence-free.

\paragraph{Vacuum energy.}

We can include
the vacuum energy contributions, which at one-loop  have the form \cite{Steinacker:2023myp}
\begin{align}
\label{vacuum-energy}
  S_{\rm vac} = -\Tr \rho^{-2} V_0(r,m_\cK)
  = -\int d^4x \rho_M \rho^{-2} V_0(r,m_\cK)
\end{align}
where $m_\cK$ is the KK scale.
The dependence on the dilaton is plausible, since  $\rho^2 = \frac{\sqrt{G}}{\rho_M}$ measures the d.o.f. per volume \eqref{rho-eplicit-cartes}. In particular, $\rho^2$ is expected to be stabilized at the quantum level.
Note that the vacuum energy density is finite due to cancellations resulting from maximally supersymmetry. 
This vacuum energy can be taken into account using \eqref{vacuum-energy-variation}
\begin{align}
     \d S_{\rm vac} &= \int d^4y \rho_M \rho^{-2} V_0 e_{\dot\a\mu} \d e^{\dot\a\mu}  
     =  \int d^4y \sqrt{G} \rho^{-6} V_0 G_{\mu\nu} e_{\dot\a}^{\ \nu} \d e^{\dot\a\mu} \ .
\end{align}
This leads to 
\begin{align}
  \label{mod-Einstein-Lambda}
\frac 1{8\pi G_N} \cG_{\mu\nu} 
= T_{\mu\nu} + T_{\mu\nu}[C] 
 - G_{\mu\nu} \tilde\Lambda
 , \qquad \tilde\Lambda = \rho^{-4} V_0 \ .
\end{align}
Hence the effective "cosmological (non-)constant" 
 comprises contributions from the induced vacuum energy $\tilde\Lambda$ as well as $\overline T_{\mu\nu}[C] \propto G_{\mu\nu}$ on the cosmological background, which should cancel up to curvature $\cG_{\mu\nu}$.

\subsection{Modified Einstein equations and non-local mirage matter}
\label{sec:lin-Einstein}

In this section, we explore the physical implications of the above modified Einstein equations for perturbations of the background in the linearized regime.
We solve \eqref{anharm-tensor-def}  as 
\begin{align}
\label{C-explicit-formula-1}
    C_{\dot\a\mu} 
    &= - \frac {r^{-4}}{g^2}\a^2 \{\tilde\Box^{-1}(e^{\dot\b}[\cF_{\dot\b\dot \a}]),x^{\mu'}\} \gamma_{\mu\mu'} \nn\\
%
\delta C_{\dot\a\mu} 
    &= - \frac {r^{-4}}{g^2}\a^2 \{\tilde\Box^{-1}\delta (e^{\dot\b}[\cF_{\dot\b\dot \a}]),x^{\mu'}\} \gamma_{\mu\mu'}  
\end{align}
where \eqref{Box-tilde-Box}
\begin{align}
\label{Box-tilde-def}
    \tilde \Box &:= r^{-4} \a^2 \{x^\mu,\gamma_{\mu\nu}\{x^\nu,.\}\} \  \approx \ \frac 13r^{-4}\a^{2} e^{\t}\Box_{G,c\to\frac 1{\sqrt{3}} c}  \ + \hs \ .
\end{align}
Note that $C_{\dot\a\mu}$ is essentially divergence-free\footnote{local normal coordinates w.r.t. $\gamma^{\mu\nu}$ or $G^{\mu\nu}$ are understood, and $\rho \approx const$ is assumed for simplicity.} $\del_\mu C_{\dot\a}^{\ \mu} \approx 0$
\eqref{div-constraint} and typically $\hs$ valued.
Now consider linearized fluctuations $\d {\bf T}_{\dot\a}$ around a flat background. Then 
\begin{align}
   \d \big(e^{\dot\b}[\cF_{\dot\b\dot \a}]\big) 
    &= \Box {\d {\bf T}}_{\dot\a}
     - e^{\dot\b}[\{{\d {\bf T}}_{\dot\b},t_{\dot\a}\}]
     + \{{\d {\bf T}}^{\dot\b},\cF_{\dot\b\dot \a}\}   \nn\\
    &=  \Box {\d {\bf T}}_{\dot\a}
     - \{e^{\dot\b}[{\d {\bf T}}_{\dot\b}],t_{\dot\a}\}
     - \{{\d {\bf T}}^{\dot\b},\{t_{\dot\b},t_{\dot\a}\}\}
     + \{{\d {\bf T}}^{\dot\b},\cF_{\dot\b\dot \a}\}  \nn\\
    &=  \Box {\d {\bf T}}_{\dot\a}
     + 2\{\cF_{\dot\a\dot \b},{\d {\bf T}}^{\dot\b}\} \nn\\
     &=: \cD^2 {\d {\bf T}}_{\dot\a} \approx  \Box {\d {\bf T}}_{\dot\a}
\end{align}
in the locally flat regime,
assuming Lorentz gauge $e^{\dot\a}[{\d {\bf T}}_{\dot\a}] = 0$.
 We then have 
\begin{align}
  \a^{-2} r^4 g^2\gamma^{\mu\mu'}  \d C_{\dot\a\mu'}  &\approx \  - \{\tilde\Box^{-1}\Box {\d {\bf T}}_{\dot\a},x^\mu\} 
  \approx -\tilde\Box^{-1}\Box \{ {\d {\bf T}}_{\dot\a},x^\mu\}
     = -\tilde\Box^{-1}\Box \d e_{\dot\a}^\mu \ .
\end{align}
 For the unperturbed background \eqref{deformed-background}, the "mirage"  energy-momentum tensor  is then
\begin{align}
\d T_{\mu\nu}[C] - \frac 12 G_{\mu\nu} \d T[C]
&=  2e^{\dot\alpha}_\mu\d C_{\dot\a\nu}  + (\mu \leftrightarrow\nu) \nn\\
 &= - \frac {2\a^{2}}{r^4 g^2} \ \gamma_{\nu'\nu}e^{\dot\alpha}_\mu\tilde\Box^{-1}\Box \d e_{\dot\a}^{\nu'} + (\mu \leftrightarrow\nu) \nn\\
 &= -\frac {2\a^{2}}{r^4 g^2}\ \gamma_{\nu'\nu} \gamma_{\mu'\mu} \tilde\Box^{-1}\Box \d \gamma^{\mu\nu }\ .
\end{align}
Raising the indices with $G^{\mu\nu} = \rho^{-2}\ \gamma^{\mu\nu}$,
we obtain 
\begin{align}
       \d T^{\mu\nu}[C] - \frac 12 G^{\mu\nu} \d T[C] 
      &\approx  - \frac {2\a^{2}}{r^4 g^2}
      \tilde\Box^{-1}\Box_G \d G^{\mu\nu }  
\end{align}
recalling $\Box \sim \rho^2 \Box_G$.
Therefore
\begin{align}
\label{T-C-expression-metric}
\boxed{\ 
    \d T^{\mu\nu}[C] 
       \approx -\frac {2\a^{2}}{r^4 g^2} \tilde\Box^{-1}\Box_G  \d \bar G^{\mu\nu} 
        }
\end{align}
where 
\begin{align}
  \d \bar G^{\mu\nu} =  \d G^{\mu\nu} - \frac 12
     G^{\mu\nu}(G_{\r\s}\d G^{\r\s}) \ 
\end{align}
is the trace-reversed metric perturbation.  $T^{\mu\nu}[C]$ is divergence-free due to the harmonic gauge condition $\del_\mu \delta\bar G^{\mu\nu} = 0$ (in local normal coordinates), which holds provided the torsion of the background is negligible as shown in appendix \ref{sec:harmonic-gauge}. 
Then the linearized modified Einstein equations \eqref{mod-Einstein} take the form  
\begin{align}
  \label{lin-mod-Einstein}
  \d \cG^{\mu\nu} 
&= 8\pi G_N (T^{\mu\nu} + \d T^{\mu\nu}[C])  \nn\\
&= 8\pi G_N \big(T^{\mu\nu} - \frac {2\a^{2}}{r^4 g^2}
\tilde\Box^{-1}\Box_G \d \bar G^{\mu\nu}\big)
\end{align}
(note that dilaton and axion would enter only at the quadratic level).
Recalling 
 $\d\cG^{\mu\nu} = -\frac 12\Box_G\d \bar G^{\mu\nu}$ in harmonic gauge\footnote{Our conventions are $G^{\mu\nu} = \bar G^{\mu\nu} + \d G^{\mu\nu}, \ \d G = G_{\mu\nu} \d G^{\mu\nu}$ and $\Box_G \approx - G^{\mu\nu}\del_\mu\del_\nu$ with mostly positive signature.}, the above equations reduce to
\begin{align}
 \boxed{\ 
  \big(1 - m^2 \tilde\Box^{-1}\big) \d \cG^{\mu\nu} 
  \approx  8\pi G_N T^{\mu\nu}
  \ }
  \label{lin-mod-Einstein-4}
\end{align}
in terms of the (Cartesian) cross-over scale\footnote{The dimensions will be straightened in \eqref{m-cross-def-2}.}
\begin{align}
\label{m-cross-def-1}
m^2 :=  \frac{32\pi\a^2}{ r^4 g^2} G_N \ .
\end{align}
This gives
\begin{align}
     \label{lin-mod-Einstein-5}
  %
\d \cG^{\mu\nu} 
  &\approx 8\pi G_N \frac{\tilde\Box}{\tilde\Box - m^2} T^{\mu\nu} 
   = 8\pi G_N (T^{\mu\nu} + \delta T^{\mu\nu}) , 
\end{align}
where 
\begin{align}
\label{deltaT-T-relation}
     \d T^{\mu\nu} := \frac{m^2}{\tilde\Box - m^2} T^{\mu\nu} \equiv \d T^{\mu\nu} [C] \ .
\end{align}
 This has the form of modified linearized Einstein equations
 \begin{align}
 \label{lin-Einstein-mod-1}
 \boxed{\ 
\d\cG^{\mu\nu}  = 8\pi G_N (T^{\mu\nu} + \delta T^{\mu\nu}) \ .
\ }
 \end{align}
We denote $\d T_{\mu\nu} \equiv  \d T_{\mu\nu} [C]$ as {\bf mirage} (dark) matter, since it is acts as an apparent matter source in the linearized Einstein equations, arising
as a "non-local reflection" of real matter due to the non-local form of the Yang-Mills action  in terms of the frame.
The IR scale $m$ characterizes the scale beyond which GR is significantly modified.
For  scales shorter than $m^{-1}$ (in Cartesian coordinates), we obtain (linearized) GR, since
$\delta T^{\mu\nu}[C] \approx 0$ for $\tilde\Box \gg m^2$. For longer distances, a screening behavior due to mirage matter is found, with $\d T_{\mu\nu}[C](k) \to - T_{\mu\nu}(k)$ in the IR limit $k\to 0$. Hence  matter is screened
for very long distances, so that linearized (!) gravity effectively has a finite range.
This modification is distinct from conventional modifications of GR such as $f(\cR)$. 
However it is similar to massive gravity in the quasi-static regime, where \eqref{Box-tilde-Box}
\begin{align}
    \tilde\Box \sim  \frac 13r^{-4}\a^{2} e^{\t}\Box_{G} \ ,
\end{align}
so that \eqref{lin-mod-Einstein-4} reduces to 
\begin{align}
    -\big(\Box_G - m^2_{\rm cross}\big) \d \bar G^{\mu\nu} 
  &\approx  16\pi G_N T^{\mu\nu}
  \label{modified-Poisson}
\end{align}
with negative\footnote{The negative sign need not imply an instability, because we have neglected all effects due to vacuum energy, which dominate very long length scales.} mass term set by the (metric) cross-over scale
\begin{align}
     \label{m-cross-def-2}
     m^2_{\rm cross} &= 3 r^{4}\a^{-2}e^{-\tau} m^2 
     = \frac 1{L_{\rm cross}^2} \ .
\end{align}
Taking into account \eqref{m-cross-def-1},
this sets a gravitational cross-over scale
\begin{align}
\label{L-cross}
     L_{\rm cross}^2 &=  \frac {e^{\tau}}{96\pi}   \frac{g^2}{L_{\rm Pl}^2}  \ 
\end{align}
Note that $g^2$ now has effectively the dimension ${\rm length}^{4}$, consistent with \eqref{LNC-a-estimate} and \eqref{Hubble-scale}. Then $L_{\rm cross}$ becomes large if $L_{\rm Pl}$ becomes small compared to $e^\tau g^2$, defining an IR scale at sufficiently late times.
It separates the Einstein-Hilbert (at  shorter scales) from the Yang-Mills regime of emergent gravity (at longer scales).
Finally,
we can recast equations \eqref{T-C-expression-metric} and \eqref{mod-Einstein} or \eqref{mod-Einstein-Lambda} as 
\begin{align}
\label{T-C-expression-metric-2}
\boxed{\ 
  \frac{1}{m^2} \tilde\Box \, T^{\mu\nu}[C] 
      \ \approx \ \frac{1}{8\pi G_N}\cG^{\mu\nu} \ = \   T^{\mu\nu} + T^{\mu\nu}[C] - G^{\mu\nu} \tilde\Lambda \   
   \   }
\end{align}
where $\tilde\Lambda = \rho^{-4} V_0$ describes the induced vacuum energy \eqref{vacuum-energy}.
Here $\d\cG^{\mu\nu}$ is replaced by $\cG^{\mu\nu}$, since the linearization should always be applicable locally in local normal coordinates \cite{Steinacker:2024unq},  hence these equations are expected to hold more generally. 

We have obtained a system of coupled tensorial  equations governing emergent gravity on spacetime $\cM^{1,3}$ via the Einstein tensor $\cG^{\mu\nu}$, the energy-momentum tensor $T^{\mu\nu}$ of matter, and
the anharmonicity tensor $T^{\mu\nu}[C]$ interpreted as "mirage matter". The latter leads to  IR modifications of GR as illustrated below. It should be clear that $T^{\mu\nu}[C]$ is a purely geometric object here, rather than an energy-momentum tensor of some actual (dark) matter.

Several simplifying assumptions were made in the above derivation: 
Extra contributions on the rhs involving $(\del\rho)^2$ and an axial vector field are dropped, which can easily be included following \cite{Kumar:2023bxg}.
The antisymmetric tensor $B_{\mu\nu}$ in \eqref{TC-B-def} was dropped, assuming that it will mainly be associated with the axionic contributions. 
The harmonic gauge condition only holds if the torsion is negligible, hence in the weak gravity regime.
Finally, it should be kept in mind that $T^{\mu\nu}[C]$ and $\cG^{\mu\nu}$ are $\hs$-valued objects,
since $\tilde\Box$ is $\hs$ valued.
This is important in the non-linear regime, and will be elaborated up to quadratic order in section \ref{sec:mode-expansion}.

\paragraph{Gravitational crossover scale.}

The IR crossover scale between the GR and YM regime is set by $m^2 = k^2$ in Cartesian coordinates, or equivalently $L^2_{\rm cross}$ \eqref{L-cross} in the effective metric.
This length scale grows with the cosmic expansion, leading to a hierarchy of scales
\begin{align}
    \frac{L^2_{\rm cross}}{L_{\rm NC,G}^2}
    &\sim \frac{e^{\tau} g^2}{L_{\rm Pl}^{2}L_{\rm NC,G}^2}
     = g_{\rm YM}^2  \frac{L_{\rm NC,G}^2}{L_{\rm Pl}^{2}}
\end{align}
using  \eqref{LNC-a-estimate}, where $g_{\rm YM}^2 = g^2 \rho^{-2}$ is the  Yang-Mills coupling.
Equivalently,
\begin{align}
    \frac{L^2_{\rm cross}}{L_{\rm UV,G}^2}
    &\sim g_{\rm YM}^2  e^{\tau}\, \frac{L_{\rm UV,G}^2}{L_{\rm Pl}^{2}} \ .
\end{align}
Hence a large separation of scales $L_{\rm cross} \gg L_{\rm NC,G} \gg  L_{\rm Pl}$ indeed arises provided $L_{\rm NC,G}^2 \gg L_{\rm Pl}^{2}$, extending the hierarchy \eqref{Hubble-scale}.
Assuming that $L_{\rm cross}$ is 10 times the galactic scale $10^{22}m$ and setting $g_{\rm YM} = 1$, the noncmmutativity scale (or $\hs$ scale) would be
$L_{\rm NC,G} \approx 5\times 10^{-7}$m, or about $1$eV. However, $L_{\rm NC,G}$ is the scale associated with the onset of the $\hs$ sector, presumably with little relevance to the low-energy physics.
The relevant UV scale for QFT should be $L_{\rm UV,G} \sim e^{-\frac 14 \tau} L_{\rm NC,G}$  \eqref{Hubble-scale}. Using $\frac{L_{\rm cosm}}{L_{\rm NC,G}} \approx O(10^{32}) \approx
e^{\frac 12\tau}$, we obtain $L^{-1}_{\rm UV,G} \approx 10^{16}$eV (or lower if $g_{\rm YM} < 1$). These scales  seem reasonable, provided a large scale $L_{\rm NC,G} \approx 10^{-7}$m is physically viable. In particular, the weak coupling regime $g_{\rm YM} < 1$ \cite{Steinacker:2026jzp} should be viable.

These considerations point to the necessity to properly understand the physical significance of the scale $L_{\rm NC,G}$.
This will strongly depend on the low-energy role of the higher-spin modes, and notably on whether or not they acquire a mass through quantum effects. An improved understanding of the model is clearly needed.


\subsection{Implications: dark (mirage) matter halo and extra modes}
\label{sec:mirage-halo}

In the static linearized regime, $T^{\mu\nu}[C]$  \eqref{T-C-expression-metric} reduces to 
\begin{align}
\label{T-C-metric-static}
     T^{\mu\nu}[C] 
      = -6 r^{-4}
      e^{-\tau} \d \bar G^{\mu\nu}  \ .
\end{align}
This is reminiscent of massive gravity, with a dynamical (negative) mass parameter that changes along the cosmic evolution.
 
\begin{figure}[ht!]
    \centering
     \includegraphics[scale=0.45]{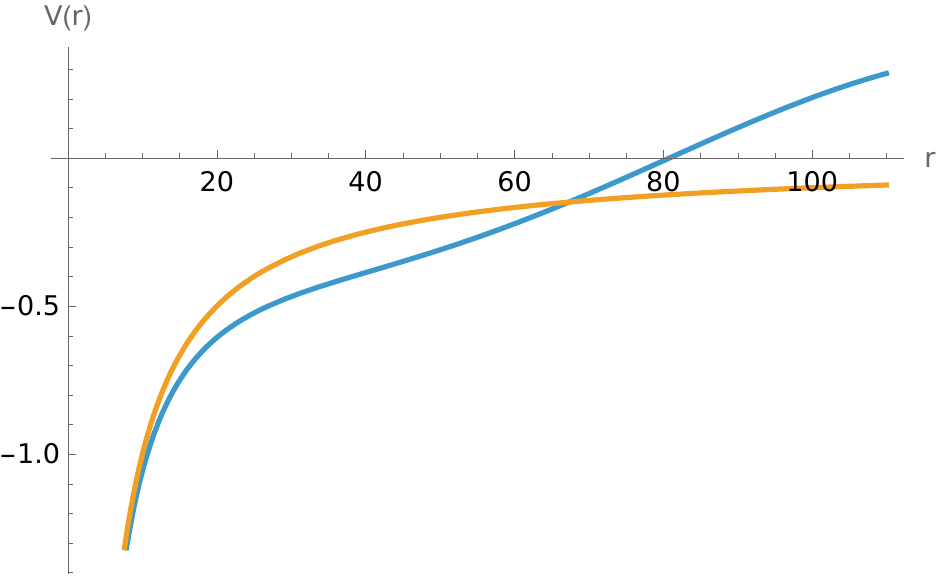}
    \ \ \ \includegraphics[scale=0.45]{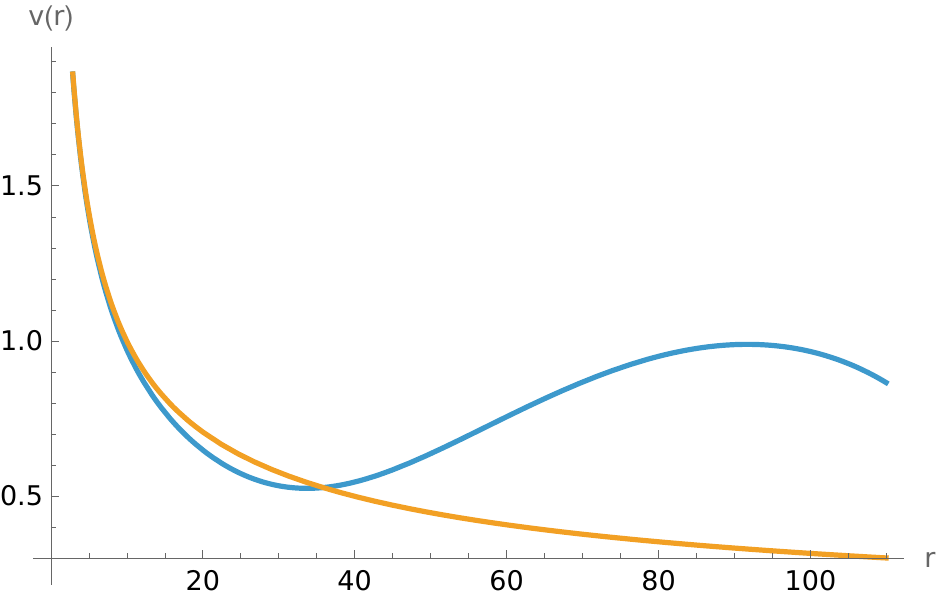}
    \caption{Gravitational potential (left) and rotation curves (right) for a point mass (yellow) and including the mirage halo (blue).}
\label{fig:rotation-velocities}
\end{figure}

To understand the effect of mirage matter more explicitly, consider a static point mass $M$ localized at the origin, with 
\begin{align}
    T_{00} = M\delta^{(3)}(r) \ .
\end{align}
Noting that
\eqref{tilde-Box-planewave-2}
\begin{align}
\label{Box-tilde-grav}
    \bar \Box \sim  \del_0^2 - \frac 13\del_i\del^i \quad (+ \hs)
\end{align} 
(in Cartesian coordinates) is proportional to the metric d'Alembertian but with speed of light reduced by a factor $\frac 1{\sqrt{3}}$, the mirage e-m tensor is given by the Greens function for the Helmholtz operator 
as
\begin{align}
 T_{00}[C] \sim  \frac{m^2}{-\frac 13\del_i\del^i - m^2 \ + \hs} M\delta^{(3)}(x)
     \sim \frac {3m^2}{4\pi r} M\cos(\sqrt{3} m r) \qquad (+ \hs) 
     \label{halo-pointmass-explicit}
\end{align}
consistent with \eqref{T-C-expression-metric}.
This amounts to some effective matter halo with small density proportional to $m^2 \ll M^2$ around the point mass $M$; note that 
$T_{00}[C]$ also has significant $\hs$-valued content. 
The equation for the linearized metric perturbation is then\footnote{recall that $G_{\mu\nu} = \bar G_{\mu\nu} - \d G_{\mu\nu}$ in our conventions.}
\begin{align}
   - \Box_G \d G_{00} = 16\pi G_N M \big(\delta^{(3)}(x) + \frac{3m^2}{4\pi r} \cos(\sqrt{3}mr)\big) \ . 
\end{align}
The halo density $T_{00}[C]\sim \frac {m^2M}r\cos(mr)$ is negligible at short distances, 
and we recover the gravitational field of linearized GR\footnote{This might be modified upon including the dilaton into the action, which  enters in the non-linear regime.}.
However at larger distances of order $L_{\rm cross}$ \eqref{L-cross}, the halo leads to flattened or increasing rotational 
velocity curves  around the central mass (and stronger gravitational lensing) in some regime, reminiscent of  dark matter.
The density $T_{00}[C]\sim \frac {\cos(mr)}r$
is somewhat similar to empirical dark matter halos such as the NFW profile \cite{Navarro:1995iw}. The oscillatory behavior should probably not be taken too seriously, but the existence of extended halos seems to be a solid feature.
The gravitational potential and the resulting rotation velocities around the point mass with and without the induced halo are plotted in figure \ref{fig:rotation-velocities}.
Of course this halo is not rigid but dynamical, determined through \eqref{deltaT-T-relation} by matter via a propagator. It naturally mimicks cold dark matter, as long as $T_{\mu\nu}$ is dominated by $T_{00}$. 
However, $T_{\mu\nu}[C]$ is of purely geometric origin here.

It is interesting to 
observe that in the quasi-static regime, the modification due to $T[C]$ are similar as in linearized gravity with {\em negative} mass term; this is already clear from \eqref{T-C-metric-static}. As such, they are
milder than in massive gravity: 
within a volume of radius $r$, it contributes an effective mirage mass  of order $O(rm)^2$:
\begin{align}
    M(r) = \int dr r^2 T[C](r)
    \sim  M (rm)^2 \qquad \mbox{for} \ \ r \ll \ m^{-1} \ .
\end{align}
For a positive mass term, this corrections would start at $O(mr)$, so that some observational bounds on the graviton mass do not apply here.
We will see below that the instability associated with negative ${m}^2$ should be cured by including $\hs$ components.
The total mass of the halo within the first peak $\sqrt{3}mr<\pi/2$ is of order $5 M$. However, the total halo mass is $\int  T_{00}[C] = -M$,
since $T_{\mu\nu}[C](q) \sim - T_{\mu\nu}(q)$ for $q\to 0$.
Then the Yang-Mills (YM) action dominates,
leading to a screening of gravity.
Hence the cosmic expansion is less sensitive to matter than in GR.
The cosmic background carries also a time-dependent $T_{\mu\nu}[C] \sim G_{\mu\nu}$ \cite{Kumar:2023bxg}, reminiscent of dark energy.

The equations \eqref{T-C-expression-metric-2} support in fact soliton-like solutions without actual matter. For example, the homogeneous solution $\frac {1}{r} \sin(m r)$ of the Helmholtz equation gives rise to a static solution for mirage matter $T_{\mu\nu}[C] \sim \frac {1}{r} \sin(m r)$, which may be detached from actual matter and acquire dynamics.

Taking into account time dependence,  we observe that 
$\Box \d \bar G_{\mu\nu} = 0$ with $T_{\mu\nu}[C] = 0$ is a vacuum solution, even for $m>0$. Hence the standard gravitational wave solutions of (linearized) GR are recovered, without any modification of the dispersion relation. This is in contrast to massive gravity \cite{deRham:2014zqa,Hinterbichler:2011tt}, which is strongly constrained from observations of gravitational waves at cosmic distances. 
Moreover, the equation \eqref{lin-mod-Einstein-4} for the metric  admits new non-Ricci-flat vacuum modes $(-\tilde\Box + m^2) \d \bar G_{\mu\nu} = 0$, which from the GR point of view look like mirage matter $T_{\mu\nu}[C]$ waves. They typically propagate with velocity $c/3$, but appear to become tachyonic (and unstable) in the extreme IR  scale $L_{\rm cross}$. This instability 
is expected to be cured by taking into account vacuum energy and $\hs$ contributions, which modifies their dispersion relation in the IR as elaborated in section \ref{sec:1-loop-modes}.

This points to an important technical caveat. In the above derivation, all $\hs$ contributions were ignored. This is justified at the linearized level, since the $\hs$ components of the background metric vanish in local normal coordinates. However they are important at the quadratic level, and contribute to the action and
to the dilaton \eqref{dilaton-vacuumenergy-A} even for transverse-traceless modes.
We will see in section \ref{sec:1-loop-modes} that this can stabilize the above modes. Moreover,
$T_{\mu\nu}[C]$ should be expected to have significant $\hs$ components due to $\tilde\Box$.

\subsection{Background solution}

The rigid background ${\bf T}^{\dot\a} = t^{\dot\a}$ \eqref{matrix-background-full} is a solution of the classical IKKT model only upon adding a mass term, which would break SUSY.
However, the generalized background ${\bf T}^{\dot\a} = \a(\tau) t^{\dot\a}$ \eqref{deformed-background} is expected to be stabilized by quantum corrections for suitable $\a(\tau)$,
since its structure is the most general compatible with the global $SO(3,1)$ symmetry up to gauge transformations.
Here we indicate how this can be seen from the 1-loop action.

To derive the equations of motion for the  background,  consider fluctuations 
$\d T^{\dot\a} = \cA^{\dot\a}$. The classical YM action contributes
\begin{align}
\label{S-YM-A}
   \d S_{\rm YM} = -4\int
 d^4y \sqrt{G}\rho^{-2} \a^3 \cA_{\dot\a} \Box_t t^{\dot\a}
  =  -12\int
 d^4y \rho_M r^{-2} \a^3 t^{\dot\a} \cA_{\dot\a} \ 
\end{align}
(neglecting derivative contributions of $\a$, cf. \eqref{locally-flat-estimate}).
The variation of the 1-loop
vacuum energy contribution\footnote{We assume that the gravitational part is sub-leading here, because it is a higher-derivative contribution.} is
\begin{align}
\label{vacuum-energy-variation}
    \d S_{\rm vac} &= -\int d^4y \rho_M \d\rho^{-2} V_0  
    = \int d^4y \rho_M \rho^{-2} V_0 \bar e_{\dot\a\mu} \d e^{\dot\a\mu} \nn\\
    &= -\int d^4y \rho_M  \{\rho^{-2}\a^{-1}e^{-\tau} V_0,x_\mu\} \cA^\mu \
   = -\int d^4y \rho_M  r^3\frac{d (\rho^{-2}\a^{-1}e^{-\tau} V)}{d x^4} t_{\dot\a} \cA^{\dot\a}
\end{align}
upon partial integration,
noting that $\d\rho^2 = \rho^2 \bar e_{\dot\a\mu} \d e^{\dot\a\mu}$ \eqref{del-rho2} and $\{x^4,x^\mu\} = r^3 t^\mu$ in Cartesian coordinates.
Recalling that $x_4 = r \sinh(\tau)$ is a cosmic time parameter,
this leads to  the equation of motion
\begin{align}
\label{cosm-eom-rho}
    \frac{d}{d\tau} (\rho^{-2}\a^{-1} e^{-\tau} V_0) =  x_4 \frac{d}{dx_4} (\rho^{-2}\a^{-1} e^{-\tau} V_0) &= - 6r^{-4} e^\tau \a^3 \ .
\end{align}
Hence the covariant background is indeed a solution for suitable $V_0$ and $\a(\tau)$, without adding any mass term to the IKKT model. A more detailed analysis is left for future work.

\section{Mode expansion and propagating modes}
\label{sec:mode-expansion}

To complement and refine the above geometric analysis, we work out in this section the mode expansion of the geometrical or gravitational sector, taking into account all $\hs$ contributions. We will mostly work in the locally flat regime using the approximations \eqref{locally-flat-estimate}. Hence the analysis cannot be fully trusted in the extreme (cosmic) IR regime.

Consider for simplicity the undeformed background ${\bf T}^{\dot\a} = t^{\dot\a}$.
Then the most general $\cC^1$ valued deformation $\TT^{\dot\a} \to  {\bf T}^{\dot\a} = t^{\dot\a} + \cA^{\dot\a}$ is given by
\begin{align}
\label{A-general-onep}
    \cA^{\dot\a} = A^{\dot\a\mu}(y) u_\mu
\end{align}
where $A^{\dot\a\mu}$ has dimension mass.
Upon gauge fixing with 
the Lorentz gauge 
\begin{align}
    0 = \{{\bf T}_{\dot\a}, \cA^{\dot\a}\}
      &=  \{t_{\dot\a}, A^{\dot\a\mu} \}u_\mu
       + A^{\dot\a\mu}  \{t_{\dot\a}, u_\mu\}  
    \approx  \{t_{\dot\a}, A^{\dot\a\mu} \}u_\mu \ ,
\end{align}
$\cA$ is governed by the equation of motion\footnote{The extra term $\{\{{\bf T}^{\dot\a},{\bf T}^{\dot\b}\},\cA_{\dot\b}\}$ can be interpreted in terms of a coupling to the (Weitzenb\"ock) torsion \eqref{Weitzenbock-torsion-def}, which we assume to be negligible in the local regime under consideration.} 
\begin{align}
  0 =  (\Box + 2\{\{{\bf T}^{\dot\a},{\bf T}^{\dot\b}\},.\})\cA_{\dot\b} \approx \Box \cA_\nu
\end{align}
in the same approximation. 
We can (or should) assume the space-like condition
\begin{align}
    A^{\dot\a 0} = 0 
\end{align}
at the reference point $\xi = (\xi^0,0,0,0)$
 as $u$ is space-like, which comprises 12  independent modes.
Upon gauge fixing
this leaves 12 - 3 = 9 dof.
The pure gauge modes take the form
\begin{align}
     \cA^{\dot\a(g)} 
     &=   \{{\bf T}^{\dot\a}, \Lambda\} 
     = \{{\bf T}^{\dot\a}, \Lambda^\mu u_\mu\} 
      \approx  \xi_4 (\del^{\dot\a} \Lambda^\mu) u_\mu \nn\\
    A^{\dot\a \mu(g)} 
     &=  \del^{\dot\a} \Lambda^\mu
\end{align}
where we can or should assume $\Lambda^0 = 0$.
As usual in Yang-Mills theory, there are residual (on-shell) gauge transformations given by 
\begin{align}
  \{{\bf T}_{\dot\a}, \cA^{\dot\a(g)}\}
  =  \{{\bf T}_{\dot\a},\del^{\dot\a} \Lambda^\mu u_\mu \}    &\approx \{{\bf T}_{\dot\a}, \del^{\dot\a} \Lambda^\mu \}u_\mu = 0 \qquad 
     \mbox{for}\quad \Box \Lambda^\mu \approx 0 \ ;
\end{align}
the precise form is given in \cite{Steinacker:2019awe}.
After subtracting these 3 residual on-shell gauge modes, we obtain $6=5+1$ physical on-shell modes. 
These are tentatively interpreted in terms of 5 would-be massive gravitons and one extra (axionic) scalar mode. 

\subsection{Physical modes}

To make the physical on-shell modes of the YM action explicit, consider plane wave modes
\begin{align}
\label{cA-modes}
  A^{\dot\a \mu}(y) = A^{\dot\a \mu} e^{i k x}
\end{align}
for 12  polarizations $A^{\dot\a \mu}$ with $A^{\dot\a 0} = 0$, subject to the Lorentz gauge
\begin{align}
\label{Lorentz-gauge-k}
    k_{\dot\a} A^{\dot\a \mu} = 0 = \del_{\dot\a} A^{\dot\a \mu} \ .
\end{align}
The on-shell condition\footnote{There will be extra on-shell modes once the Einstein-Hilbert term is included.} is $0 = k^2 \ = k_{\mu} k^{\mu}$.
Assuming that 
\begin{align}
\label{k-null-explicit}
    k^{\dot\a} = (k,k,0,0), \qquad i.e. \  k^- = 0
\end{align}
is light-like along the $\mu = +$ direction,
this leaves 
the following modes 
\begin{align}
A^{2 \mu}, \  A^{3 \mu}, \  A^{+ \mu}
\end{align}
for space-like $\mu$.
This contains the apparent null modes $A^{+ \mu}$, 
but these are precisely the 3 residual  on-shell  pure gauge modes 
\begin{align}
    \cA^{\dot\a \mu(g)} 
     &=  k^{\dot\a} \Lambda^\mu e^{i k y} \qquad 
     \mbox{for}\quad  k^2 = 0
\end{align}
(for space-like $\mu)$
which are factored out from the physical Hilbert space, as usual in YM.
Therefore all remaining 6 modes are physical with positive norm\footnote{These 6 physical modes were obtained in a different way in \cite{Steinacker:2019awe,Sperling:2019xar} using the global geometry.  5 modes were identified as physical gravity modes 
    $\cA^{\dot\mu} = \{x^{\dot\mu},\phi^{(2)}\}$.
Among these, the massless (transversal traceless) graviton modes are obtained from $\phi^{(2,0)}$, i.e. $v^{ij} k_i = 0$.}
, given by 
\begin{align}
\cH_{phys} = \{A^{2 \mu}, \  A^{3 \mu} \}
\end{align}
for space-like $\mu$.
These are automatically orthogonal to pure gauge modes due to \eqref{Lorentz-gauge-k}, and they satisfy the space-like condition
\begin{align}
\label{spacelike-A}
    \xi_{\dot\a} A^{\dot\a \mu} = 0 \ .
\end{align}
We can separate these into {\bf transversal modes $A_{(tran)}^{\dot\a \mu} k_\mu = 0$}
and {\bf longitudinal modes $A_{(long)}^{\dot\a 1}$} along $\mu=1$ which are orthogonal to the transversal modes; these correspond to  the extra $\mu=+$ modes \eqref{physical-e-modes}.

\paragraph{Frame modes.}

Recall that the $A$ modes are {\bf potentials} for the frame.
The effective frame modes are 
obtained as 
\begin{align}
\label{frame-modes}
    \d e^{\dot\a \nu} &= \{A^{\dot\a \mu}u_\mu,y^\nu\} 
    =  \{A^{\dot\a \mu},y^\nu\}u_\mu +  A^{\dot\a \mu} \{u_\mu,y^\nu\}  \nn\\
    &=  \{A^{\dot\a \mu},y^\nu\}u_\mu +  A^{\dot\a \mu} r (\delta^\nu_\mu - u^\nu u_\mu)
  \nn\\
    &\approx   \{A^{\dot\a \mu},y^\nu\}u_\mu 
     = i A^{\dot\a \mu} u_\mu k_\sigma \theta^{\sigma\nu}  \nn\\
     &= i r A^{\dot\a \mu} u_\mu k_\sigma (\xi^\sigma u^\nu - \xi^\nu u^\sigma) 
\end{align}
(we will drop the $i$ from the plane wave factors henceforth),
which is divergence-free
\begin{align}
\label{e-div-free}
\del_\nu \d e^{\dot\a \nu}  
 =  k_\nu \d e^{\dot\a \nu}  
 =  A^{\dot\a \mu} u_\mu k_\sigma \theta^{\sigma\nu}k_\nu = 0 \ 
\end{align}
in Cartesian coordinates.
Note that they are no longer space-like in the covariant index $\nu$.
The propagating  frame modes contain only transversal space-like $\dot\a$ since 
\begin{align}
\label{e-div-frame}
\del_{\dot\a} \d e^{\dot\a \nu}   =
k_{\dot\a} \d e^{\dot\a \nu}   &=  k_{\dot\a}  A^{\dot\a \mu} u_\mu k_\sigma\theta^{\sigma\nu} = 0
\end{align}
due to the Lorentz gauge condition \eqref{Lorentz-gauge-k}, and the space-like condition 
\eqref{spacelike-A}
\begin{align}
\label{spacelike-e}
    \xi_{\dot\a} \d e^{\dot\a \mu} = 0 \ .
\end{align}
The three pure gauge  modes are
\begin{align}
\label{pure-gauge-v-def}
    \d e^{\dot\a \nu (g)} = k^{\dot\a}\big(k_0   u^\nu 
    - \delta^{\nu,0}  k_\sigma u^\sigma \big) \Lambda^\mu u_\mu 
    =: k^{\dot\a} v^\nu
\end{align}
(with $v^\nu k_\nu = 0$), which are factored out from $\cH_{phys}$. 
Hence for fixed $k^\mu$ of the form \eqref{k-null-explicit},
there are 6 physical frame modes given by 
\begin{align}
\{\d e^{2 \mu}, \  \d e^{3 \mu} \}, \qquad \mbox{for}\ \  \mu = +,2,3 \ .
\label{physical-e-modes}
\end{align}
Their projection on the classical sector $\cC^0$ is obtained using \eqref{u-average}
\begin{align}
\label{frame-classical-A}
 [\d e^{\dot\a \nu}]_0 &= r A^{\dot\a \mu} 
  k_\sigma[(\xi^\sigma u^\nu - \xi^\nu u^\sigma)u_\mu ]_0 
    = \frac{r^2}3 \sinh\t \big(k_0 A^{\dot\a \nu} 
    - \delta^\nu_0  A^{\dot\a \sigma} k_\sigma  \big) \ .
\end{align}
The two $\nu=+$ modes arising from $A_{(long)}^{\dot\a 1}$
are longitudinal $[\d e^{\dot\a \nu}_{(long)}]_0 \propto w^{\dot\a} k^\nu$, while the 
$A_{(tran)}^{\dot\a\nu}$ lead to 4 transversal frame modes 
with $[\d e^{\dot\a 0}_{(tran)}]_0 = 0$ since $A^{\dot\a 0} = 0$. 

It is worth pointing out that 
the trace 
\begin{align}
    \eta_{\dot\a\nu} \d e^{\dot\a \nu}
      \propto (k_\sigma \xi^\sigma) A^{\dot\a \mu}   u_{\dot\a} u_\mu
\end{align}
using \eqref{spacelike-A} reduces to $A^{\dot\a \mu}\delta_{\dot\a\mu}$ upon projection to $\cC^0$, but involves all modes at the $\hs$ level.

\paragraph{Metric modes.}

The above physical frame modes lead to the following modes for the (rescaled) effective metric 
\begin{align}
 h^{\mu\nu}  :=  \d \gamma^{\mu\nu}  &= \sinh\t \big(\d e^{\nu \mu} +  \d e^{\mu\nu}\big)
\end{align}
while $\bar e^{\dot\a \nu} = \sinh(\t) \eta^{\dot\a \nu}$
and $\bar\gamma^{\mu\nu} = \eta_{\dot\a\dot\b} \bar e^{\dot\a \mu} \bar e^{\dot\b \nu} $
denotes the background.
Using \eqref{e-div-frame} and \eqref{e-div-free}, the physical modes satisfy the 
gauge condition
\begin{align}
    \del_\mu h^{\mu\nu}  = 0
\end{align}
in Cartesian coordinates.
This comprises 6 physical  metric fluctuations $\{h^{ab}, a,b = +,2,3\}$.
Among these, the two longitudinal physical modes  $\d e^{i+}_{(long)}$ modes \eqref{physical-e-modes}  lead to  diffeo-like modes
\begin{align}
    [h^{\mu \nu}_{(long)}]_0 = k^{\mu} w_i^\nu + 
    k^{\nu} w_i^{\mu} \ 
    \label{diffeo-modes-h-extra}
\end{align}
for $i=2,3$.
This requires some explanation. We have already factored out  from $\cH_{phys}$
the volume-preserving diffeo modes $\d e^{+\mu} \sim k^{\dot\a} v^\mu$, which also give rise to 
diffeo modes 
\begin{align}
    h^{\mu \nu (g)} = k^{\mu} v^\nu + 
    k^{\nu} v^{\mu} \ .
    \label{diffeo-modes}
\end{align}
Nevertheless,  $\cH_{phys}$ does contain the above two longitudinal
modes $e^{i+}_{(long)}$ 
whose classical part leads diffeo-like modes in the metric. 
They could be viewed as "would-be massive" graviton modes, but they also
contribute to the 
antisymmetric frame sector, which is physical here because there is no local Lorentz gauge invariance.
Note also that
there are three physical transversal metric fluctuations $\{h^{22},h^{33},h^{23}\}$ which include trace fluctuations, as the gauge group consists of volume-preserving diffeos only.
This trace mode is captured by the dilaton:


\paragraph{Dilaton modes.}

The fluctuation modes $A$ also affect the dilaton as
\begin{align}
\rho^2 &= \rho_M \det e^{\dot\a\mu} 
 = \bar\rho^2 \det(\delta^\mu_\nu +  \bar e_{\dot\a\nu} \d e^{\dot\a\mu} ) \ . 
\end{align}
This
contributes the following terms \eqref{rho-2-expand-A} to the vacuum energy $S_{vac} = -\int\rho_M \rho^{-2} V_0$  
\begin{align}
\label{dilaton-vacuumenergy-A}
\bar\rho^{2}[\rho^{-2}]_0
&= 1 - \frac{r^2 k_0}{3\a}  \tr(A)
+ \frac {(r^2 k_0)^2}{15\a^2}
     \big(\tr(AA) + \tr(AA^T) + (\tr(A))^2 \big) \ + O(A^3) \ .
\end{align}

\paragraph{Axionic modes.}

Consider finally
the axionic vector field \eqref{Weitzenbock-torsion-def}:
\begin{align}
 \d \tilde T_\gamma \sim  \varepsilon_{\dot\a\dot\b\mu\gamma} \d T^{\dot\a\dot\b\mu} 
 & \sim \varepsilon_{\dot\a\dot\b\mu\gamma}
     k^{\dot\b} \d e^{\dot\a\mu} 
\end{align}
(schematically).
This encodes the antisymmetric frame mode, and is thereby identified as gravitational axion.
Due to 
\begin{align}
\label{torsion-div}
  k_\mu \d T^{\dot\a\dot\b\mu} =
  k_{\dot\a} \d T^{\dot\a\dot\b\mu} = k_{\dot\b} \d T^{\dot\a\dot\b\mu}   = 0 \ 
\end{align}
for physical on-shell modes $k_\mu k^\mu = 0$,
it follows that
\begin{align}
    \d \tilde T_\gamma 
 \sim k_{\gamma} \varepsilon_{01ij} \d e^{ij} \ ,
\end{align}
consistent with $\tilde T_\gamma = \rho^{-2}\del_\gamma \tilde\rho$ \cite{Fredenhagen:2021bnw}. 
Note that fermions couple to the  axion, cf. \cite{Battista:2022vvl}.

\subsection{Quadratic 1-loop action and modes}
\label{sec:1-loop-modes}

Now we study the mode expansion of the combined Yang-Mills and gravitational action up to quadratic order in the fluctuations $\cA$.
We will work out the 1-loop effective action for fluctuation modes to quadratic order, taking into account their $\hs$ components.

\paragraph{$U(1)$ field strength and action.}

Consider the matrix model action \eqref{S-YM}.
Adding $\cC^1$ fluctuations to the background $\TT^{\dot\a} = \a  t^{\dot \a} + A^{\dot\a \mu}u_\mu$,
the  $\hs$ valued field strength $\cF_{\dot\a\dot \b}$ becomes
\begin{align}
\label{F-A-explicit}
  \cF^{\dot\a\dot\b} &=  \a (- \{t^{\dot\a},A^{\dot\b \mu}u_\mu\}
   + \{t^{\dot\b},A^{\dot\a \mu}u_\mu\})  \ 
    -\{A^{\dot\a\nu} u_\nu,A^{\dot\b \mu}u_\mu\} \nn\\
    &= \a e^\t (- \del^{\dot\a} A^{\dot\b \mu}
   + \del^{\dot\b} A^{\dot\a \mu}) u_\mu \ \ 
    + O(L^2_{NC} \del A\del A)
 \end{align}
dropping non-derivative terms \eqref{mass-term-YM} and the background field strength, and treating $\a e^\tau$ as constant.
Assuming  Lorentz gauge $\del^{\dot\a} A_{\dot\a} = 0$ \eqref{Lorentz-gauge-k},
this gives the quadratic contribution
\begin{align}
     \cF^{\dot\a\dot\b} \cF_{\dot\a\dot\b} &= \a^2 e^{2\t} (- \del^{\dot\a} A^{\dot\b \mu}
   + \del^{\dot\b} A^{\dot\a \mu})
 (- \del_{\dot\a} A_{\dot\b}^{\ \nu}
   + \del_{\dot\b} A_{\dot\a}^{\ \nu}) u_\mu u_\nu \nn\\
   &=- 2\a^2 e^{2\t}
 A^{\dot\b \mu}\del^{\dot\a}\del_{\dot\a}  A_{\dot\b}^{\ \nu}
   u_\mu u_\nu \quad + \del(...)\nn\\
   &= 2 A^{\dot\b \mu}\Box A_{\dot\b}^{\ \nu}
   u_\mu u_\nu   \quad + \del(...)
   \label{YM-action-quad}
\end{align}
 to the Yang-Mills action, using $\Box = -\a^2 e^{2\t} \del^{\dot\a} \del_{\dot\a}$.
Averaging over $S^2$  via $[u^i u^j] = \frac 1{3}\delta^{ij}$ gives
the expected ghost-free action for the physical fluctuations
\begin{align}
 [\cF^{\dot\a\dot\b} \cF_{\dot\a\dot\b}]_0  
 &= \frac 2{3}  A^{\dot\b i} \Box A_{\dot\b}^{\ i} \quad + \del(...) \ .
\end{align}
In the extreme IR this should be supplemented by extra terms \eqref{mass-term-YM}.

\paragraph{Torsion.}

To describe the 1-loop action, it is useful to consider the Weitzenb\"ock torsion associated to the frame $e^{\dot\a\mu}$ \eqref{Weitzenbock-torsion-def}. 
The fluctuation modes lead to 
linearized torsion modes 
\begin{align}
  \d T^{\dot\a\dot\b\mu} &= \bar e^{\dot\b}[\d e^{\dot\a\mu}] - \bar e^{\dot\a}[\d e^{\dot\b\mu}] 
= \{\cF^{\dot\a\dot\b},x^\mu\}
 \sim \theta^{\sigma\mu}\del_\sigma\cF^{\dot\a\dot\b}  
 \label{torsion-explicit-0}
\end{align}
or more explicitly
 \begin{align}
  \d T^{\dot\a\dot\b\mu} 
   &= \a e^{\t} (\del^{\dot\b} \d e^{\dot\a\mu} - \del^{\dot\a} \d e^{\dot\b\mu}) \nn\\
  &\sim r^2 \a e^{2\t}\big(\del^{\dot\b} \big(\del_0 A^{\dot\a \kappa} u^\mu u_\kappa 
    - \delta^{\mu,0} \del_\sigma A^{\dot\a \kappa}  u^\sigma  u_\kappa \big) 
    - (\dot\a \leftrightarrow\dot\b) \big) 
  \qquad \in \ \cC^0 + \cC^2
 \label{torsion-explicit}
\end{align}
using \eqref{frame-modes},
with classical part
\begin{align}
  [\d T^{\dot\a\dot\b\mu}]_0 
   &= \a e^{\t} \big(\del^{\dot\b} [\d e^{\dot\a \mu}]_0 
    - \del^{\dot\a} [\d e^{\dot\b \mu}]_0 \big)  \nn\\
  &\sim \frac {r^2}3 \a e^{2\t}  \big(\del^{\dot\b} \big(\del_0 A^{\dot\a \mu} 
    - \delta^{\mu,0} \del_\sigma A^{\dot\a \sigma}   \big) 
    - (\dot\a \leftrightarrow\dot\b) \big)  \ .
\end{align}
We can now 
 work out the induced gravity action  \eqref{S-grav-1loop} for the physical modes using
$\d T^{\dot\a\dot\b\mu} = \del_\sigma\cF^{\dot\a\dot\b} \theta^{\sigma\mu}$ \eqref{torsion-explicit-0}.
Contracting the above using \eqref{YM-action-quad} and \eqref{e-div-frame} gives
\begin{align}
   \d T^{\dot\a\dot\b\mu}
    \d T_{\dot\a\dot\b}^{\ \ \mu} \gamma_{\mu\nu}
       &= 2 \d e^{\dot\b\mu }\Box \d e_{\dot\b}^{\ \nu} \gamma_{\mu\nu} \quad + \del(...)
     \nn\\
     &=  \del_\sigma\cF^{\dot\a\dot\b}  \del_\kappa\cF_{\dot\a\dot\b}  \theta^{\sigma\mu} \theta^{\kappa\nu} \gamma_{\mu\nu} \nn\\
       &= 2 \del_\sigma A^{\dot\b \mu} u_\mu
       \Box \del_\kappa A_{\dot\b}^{\ \nu}u_\nu
        \theta^{\sigma\mu} \theta^{\kappa\nu} \gamma_{\mu\nu} \quad + \del(...) \ .
    \label{gravity-action-frame}
\end{align}
Hence the induced gravity action quadratic in the effective frame fluctuations
\begin{align}
\label{eff-frame}
    \d E^{\dot\b\mu } = \rho^{-1} \d e^{\dot\b\mu }
\end{align}
takes the familiar form
\begin{align}
\label{S-grav-1loop-eff}
    S_{\rm grav} = -  \int
 \frac{ d^4x \sqrt{G}}{16\pi G_N} \rho^{-2}\, 
 \d e^{\dot\b\mu }\Box \d e_{\dot\b}^{\ \nu} \gamma_{\mu\nu}
 = -  \int
 \frac{ d^4x \sqrt{G}}{16\pi G_N} \, 
  \d E^{\dot\b\mu }\Box_G \d E_{\dot\b}^{\ \nu} G_{\mu\nu}
\end{align}
recalling $\Box_G =\rho^{-2} \Box$ and assuming $\rho \approx const$.
This can be made more explicit in term of the $A$ modes,
using $\bar\gamma^{\mu\nu} = \a^{2} e^{2\t} \eta^{\mu\nu}$
(in Cartesian coordinates)
and  \eqref{theta-munu-explicit}
\begin{align}
   r^{-2} \theta^{\sigma\mu} \theta^{\kappa\nu} \eta_{\mu\nu} &= - (\xi^0)^2 u^\sigma u^\kappa + \xi^\sigma \xi^\kappa \qquad \in \ \cC^0 + \cC^2 \ .
\end{align}
Then
\begin{align}
 \d T^{\dot\a\dot\b\mu} \d T_{\dot\a\dot\b\mu}
     &=  2 r^2\a^{-2} e^{-2\tau} \del_\sigma A^{\dot\b \mu}\Box \del_\kappa A_{\dot\b}^{\ \nu}
   (u_\mu u_\nu)
     \big( - (\xi^0)^2 u^\sigma u^\kappa + \xi^\sigma \xi^\kappa \big)  \nn\\
    [\d T^{\dot\a\dot\b\mu} \d T_{\dot\a\dot\b\mu}]_0
    &=  -2 r^4\a^{-2}  A^{\dot\b \mu}\Box \del_\sigma\del_\kappa A_{\dot\b}^{\ \nu}
     \big[- u_\mu u_\nu u^\sigma u^\kappa 
     + u_\mu u_\nu \delta^{\sigma,0} \delta^{\kappa,0} \big]_0 \quad + \del(...) \nn\\
    &= -\frac{2}3 r^4\a^{-2} A^{\dot\b i} \Box 
    \big(\delta_{ij}
    (\del_0^2  - \frac {1}{5} \vec \del^2) 
     -  \frac {2}{5} \del_i \del_j\big)  A_{\dot\b}^{\ j} \quad + \del(...) 
\label{TT-explicit}  
\end{align}
where $\vec \del^2 = \del_i \del_j\d^{ij}$.
The projection to $\cC^0$ is evaluated using  
\begin{align}
    [u^\mu u^\nu u^\rho u^\sigma]_0 = 
    \frac 35([u^\mu u^\nu]_0 [u^\rho u^\sigma]_0 + [u^\mu u^\rho]_0[t^\nu u^\sigma]_0
    +[u^\mu u^\sigma]_0 [u^\rho u^\nu]_0 ) \ .
\end{align}
For these propagating modes, this separates into transversal and longitudinal contributions
\begin{align}
    [\d T_{(tran)}^{\dot\a\dot\b\mu} \d T^{(tran)}_{\dot\a\dot\b\mu}]_0
     &= -\frac{2}3 r^4\a^{-2} A_{(tran)}^{\dot\b i} \Box \big(\del_0^2  - \frac {1}{5} \vec \del^2\big) A_{(tran)\dot\b i}
     \nn\\
  [\d T_{(long)}^{\dot\a\dot\b\mu} \d T^{(long)}_{\dot\a\dot\b\mu}]_0
     &= - \frac{2}3 r^4\a^{-2} A_{(long)}^{\dot\b i}\Box \big(\del_0^2  - \frac {3}{5} \vec \del^2\big)A_{(long)\dot\b i} \ .
\end{align}

\paragraph{Combined action.}

Now we combine the 1-loop action \eqref{S-grav-1loop}
with the YM action \eqref{S-YM}:
\begin{align}
    \label{S-grav-full}
    S_{\rm grav+YM} &= -\int
 d^4y \sqrt{G} \Big(\frac 1{16\pi G_N} \rho^{-2} \gamma_{\mu\nu} \d e^{\dot\b\mu } \Box \d e_{\dot\b}^{\ \nu} 
  + \frac{2}{3g^2} \rho^{-2} 
   A^{\dot\b i}\Box A_{\dot\b i} \Big)  \nn\\
   &= \int d^4x \sqrt{G}\,
 \frac{r^4}3\frac{\a^{-2}\rho^{-2}}{16\pi G_N}
  \Big( A_{(tran)}^{\dot\b i}\Box  \big(\del_0^2  - \frac {1}{5} \vec \del^2\big) A_{(tran)\dot\b i} 
     +  A_{(long)}^{\dot\b i}\Box   \big(\del_0^2 - \frac {3}{5} \vec \del^2\big)A_{(long)\dot\b i}
       \nn\\
 &\qquad \quad - \frac{2}{3g^2} 
   A^{\dot\b i}\Box  A_{\dot\b i} \Big) \nn\\
  &= \int d^4x \sqrt{G}\,
 \frac{1}3\frac{1}{16\pi G_N}\,
 \rho^{-2} \a^{-2}r^4\Big( 
 A_{(tran)}^{\dot\b i} \Box \big(\del_0^2  - \frac {1}{5} \vec \del^2 - m^2 \big) A_{(tran)\dot\b i}
 \nn\\
&\qquad \qquad\qquad \qquad
+ A_{(long)}^{\dot\b i} \Box \big(\del_0^2  - \frac {3}{5} \vec \del^2 - m^2 \big) A_{(long)\dot\b i}
 \Big) \ .
\end{align}
The common factor $\Box$ implies that the transversal and longitudinal gravitational waves remain massless\footnote{note that longitudinal modes do not couple to the conserved energy-momentum tensor.}.
Here
\begin{align}
\label{m-cross-def-1a}
       m^2 = 32\pi G_N \frac{\a^2}{r^{4}g^2}
\end{align} 
recovers the (Cartesian) IR scale \eqref{m-cross-def-1} setting the boundary between the YM regime governing the cosmological evolution, and the Einstein-Hilbert regime with GR-like dynamics.
We expect that the first term dominates for perturbations with 
$\vec k^2 \gg m^2$, leading to (extended) GR. The second (YM) term dominates
in the IR regime $\vec k^2 \ll m^2$, which is the cosmic regime.

Higher-loop contributions are expected to be UV finite as well, with effective UV cutoff given by $\Lambda_{\rm NC}$ or the highest KK (SUSY breaking) scale, which is related to the Planck scale. Hence they are suppressed at low energies.

\paragraph{Dynamical modes and potentially unstable modes.}

Now we discuss solutions of the vacuum equations of motion resulting from the above quadratic action \eqref{S-grav-full}.

\begin{figure}[h!]
    \centering
\includegraphics[width=0.6\linewidth]{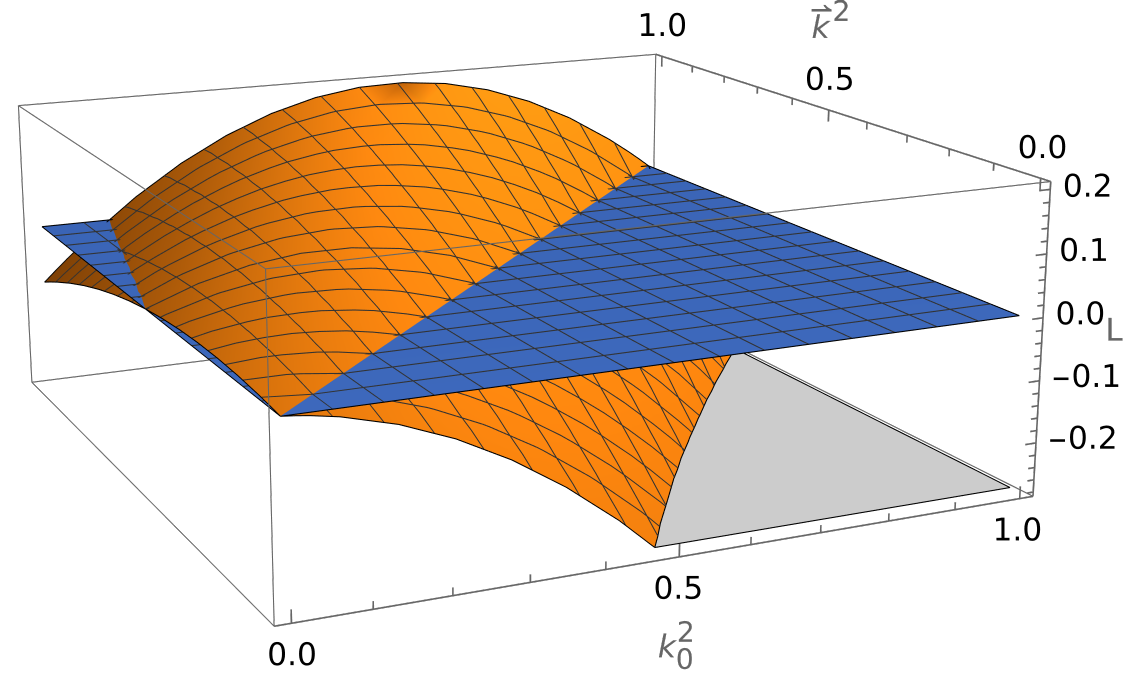}
    \caption{Kinetic action and unstable modes for $A(-k_0^2 + \vec k^2) (k_0^2 - \vec k^2/5 + 0.1)A$.}
    \label{fig:mod-disp-unstab}
\end{figure}

\begin{figure}[h!]
    \centering
\includegraphics[width=0.6\linewidth]{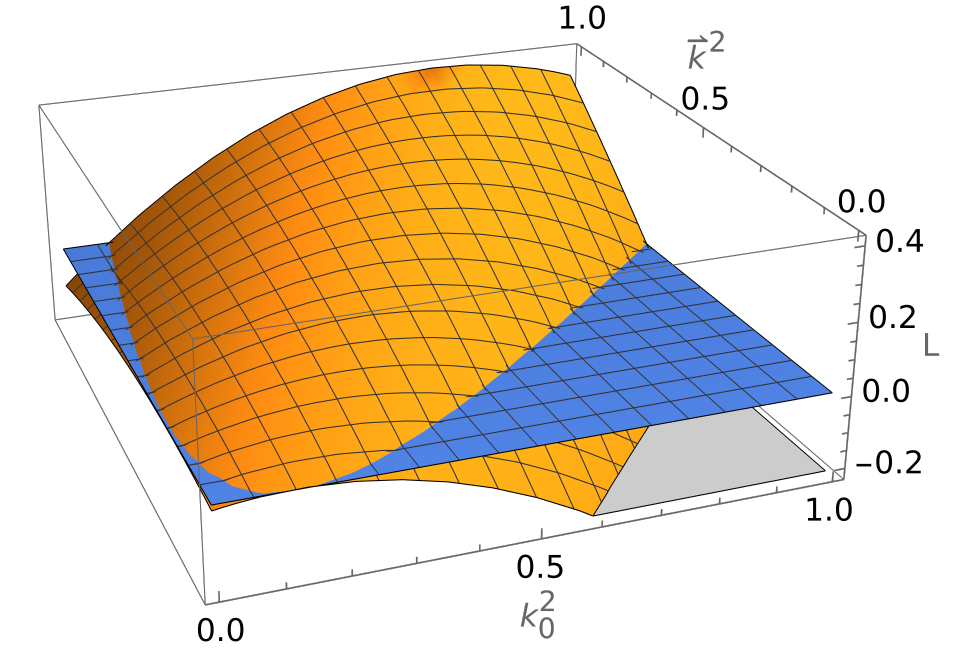}
    \caption{Kinetic action for stabilized modes $A\big((-k_0^2 + \vec k^2) (k_0^2 - \vec k^2/5 + 0.1) - 0.015 + 0.355 k_0^2\big)A\ $.}
    \label{fig:mod-disp-stab}
\end{figure}

The first observation is that 
$\Box A = 0$ and hence 
$\Box \d G^{\mu\nu} = 0$ is a solution even for $m\neq 0$, consistent with the results in Section \ref{sec:mod-einstein}. 
However, the factors $(k_0^2  - \frac {1}{5} \vec k^2 + m^2)$ etc. in \eqref{S-grav-full} lead to extra modes, which arise from the interplay between YM and E-H terms.  This is essentially consistent with \eqref{Box-tilde-grav} and \eqref{lin-mod-Einstein-5}; the different factors $\frac 13$ vs. $\frac 15$ indicates that the geometric treatment in section \ref{sec:mod-einstein} does not fully capture the $\hs$ sector.
These extra modes seem to
be tachyonic in the IR for  $\vec k^2 < m^2$, as 
can be seen in the plot of the kinetic term $A(-k_0^2 + \vec k^2) (k_0^2 - \vec k^2/5 + 0.1)A$ in figure \ref{fig:mod-disp-unstab}.
This instability can be cured by taking into account vacuum energy
contributions \eqref{dilaton-vacuumenergy-A} and the non-derivative contribution \eqref{mass-term-YM} from the Yang-Mills term. Then the quadratic effective action \eqref{S-grav-full} takes the form
\begin{align}
    S_{\rm grav+YM+vac} &= -\int\rho_M \rho^{-2} A^T \big((-k_0^2 + \vec k^2) (k_0^2 - \frac 15 \vec k^2 + m^2) + V_{00} + V_{01} k_0^2  + ... \big)A  \ .
    \label{S-grav-quad-full}
\end{align}
Note that $V_{00}<0$ arises from the non-derivative contributions \eqref{rho-2-expand-A} to the YM term, while $V_{01} > 0$ arises from vacuum energy. These terms modify the dispersion relation in the IR, and
the tachyonic instabilities can then be avoided, as illustrated in figure \ref{fig:mod-disp-stab}.
Hence the unstable new IR modes found in the linearized analysis can be stabilized via the $\hs$ components at the quadratic order. 
Matter will still generate a halo as in section \ref{sec:mirage-halo}, and (almost-) static mirage modes  with vanishing group velocity persist.
The extra propagating modes have a non-relativistic dispersion relation with group velocity less than $c$  (except at cosmic scales), assuming the above extra terms.
There is no issue with ghost-like signatures in the action, since the extra modes arises only after quantization, and only affect the background geometry.


\section{Discussion}

We derived modified Einstein equations \eqref{T-C-expression-metric-2} governing the effective metric in the IKKT matrix model on covariant quantum spacetime, obtained by combining the (semi-)classical action
with the Einstein-Hilbert action induced at one loop. 
The crucial new ingredient is a  "mirage" or anharmonicity tensor $T_{\mu\nu}[C]$ on the rhs of the Einstein equations, which arises by rewriting the Yang-Mills-type action in terms of a non-local expression for the frame. 
It describes new physics at large scales beyond $L_{\rm cross}$, which separates the length scales governed by GR and the Yang-Mills action, respectively. 

On a technical level, the description for gravity is quite non-standard: 
The  matrix background serves as {\em potential} for the frame. 
The classical Yang-Mills action cannot be expressed as a local action in terms of the frame, but only as a non-local action.
Combined with the Einstein-Hilbert action, this leads to the novel mechanism for mirage matter: The higher-order equations of motion imply extra degrees of freedom with non-local features if expressed in terms of the frame or metric. Nevertheless, standard GR and local physics is recovered at scales shorter than $L_{\rm cross}$. 
Determining this new scale requires a better understanding of the background including the internal space $\cK$.

Mirage matter 
$T_{\mu\nu}[C]$ manifests itself as a halo-like effective energy-momentum tensor which is induced around localized matter. This is reminiscent of dark matter, leading to flattened rotation curves in suitable regimes. It also has implications at cosmic scales reminiscent of some sort of dark energy.
Moreover, $T_{\mu\nu}[C]$ is governed by a modified pseudo-relativistic dynamics, and may propagate without matter. 
An instability appears in the linearized approximation in the extreme IR, which is expected to be an artifact; a stabilization mechanism by taking into account the higher-spin contributions at quadratic order is exhibited.

The specific structure of the quantum spacetime is not important in the derivation of the modified Einstein equations, it only enters the non-standard d'Alembertian $\tilde\Box$ \eqref{Box-tilde-def} governing the extra modes. Hence the results should apply quite generally to non-commutative backgrounds in the matrix model,
as long as the action is well-defined. This is not the case for commutative backgrounds where the action is UV-divergent \cite{Ho:2025htr}.

The analysis in this paper and in particular the equations \eqref{T-C-expression-metric-2} are simplified for the sake of transparency: the axionic field  $\tilde T^\mu$ is dropped, the dilaton $\rho$ is assumed to be constant, and vacuum energy is treated very superficially.
In particular,
the equations are not yet predictive at the cosmic scale; a more complete treatment along \eqref{cosm-eom-rho} requires a more detailed understanding of vacuum energy and of the internal space $\cK$. This should also settle the instability issue in the IR.
 Axions and dilatons can easily be taken into account using the results in \cite{Kumar:2023bxg}. This  might also affect to some extent the gravitational field around point masses, since $\rho$ encodes the determinant of the metric via \eqref{dilaton-def}; a more refined analysis is in order here. Nevertheless, the new physics arising at $L_{\rm cross}$ and in particular the formation of dynamical halos around matter  is expected to be a solid prediction of the matrix model on (covariant) quantum spacetime.

The appearance of higher-spin modes on 
covariant quantum spacetime may seem an undesirable complication, and the physical relevance of these modes with the relevant scale $L_{\rm NC,G}$ remains to be understood.
On the other hand, it might hold the key to resolve the singularity problem at the center of black holes: if the curvature becomes too large, the $\hs$ sector on the 6-dimensional bundle is expected to couple strongly to matter, and open up a channel to accommodate the information of in-falling matter. Specifically, the internal $S^2$ and the radial $S^2$ might mix and change roles; but this is just speculation for now.

In any case, it is clear that
the resulting gravity theory shares basic properties of GR, but goes beyond it in many ways. It provides a quantum theory of gravity, which naturally leads to features reminiscent of dark matter and dark energy. More work is required to see whether this gravity theory is physically viable.


\section*{Acknowledgement}

This paper grew out of a collaboration with Pei-Ming Ho and Hikaru Kawai, to whom I would like to express my gratitude for extensive discussions and valuable input. I would also like to thank Chong-Sun Chu, Alessandro Manta and Kaushlendra Kumar for dicsussions and related collaborations.
This work is supported by the Austrian Science Fund (FWF) grant
P36479.


\appendix

\section{Appendix}

\subsection{Local normal coordinates}
\label{sec:LNC}

We summarize some results in section 10.2.5 of \cite{Steinacker:2024unq}, which establish the crucial fact that any given $\hs$-valued matrix background
defines locally some  ordinary 3+1 dimensional geometry.
The basic observation is that
the frame ${e}_{\dot\a}^{\ \mu} = \{{\bf T}_{\dot\a},x^\mu\}$ and all derived objects (such as metric and torsion) are tensors,
considering the frame indices ${\dot\a}$ as fixed.
This means that they transform as $\hs$-valued tensors under coordinate redefinitions
\begin{align}
 \tilde y^\mu = \phi^\mu(x) \ ,
\end{align}
i.e.
\begin{align}
 {\tilde e}_{\dot\a}^{\ \mu} = \{{\bf T}_{\dot\a},\tilde y^\mu\}
 =  \frac{\del \phi^\mu}{\del x^\nu} \, {e}_{\dot\a}^{\ \nu}\,
\end{align}
assuming that $\phi^\mu(x)$ depends only on $x^\mu$.
We can take advantage of this freedom to
cast any $\hs$-valued frame locally into standard form in terms of suitably adapted
 $\hs$-valued coordinates on spacetime
 \begin{align}
 \tilde y^\mu = \phi^\mu(x,u) \ .
\end{align}
Then the dependence of the frame on $u$
can always be eliminated locally,
in analogy to Riemann normal coordinates in GR:

\paragraph{Lemma:}

Let  $e_{\dot\a} = \{{\bf T}_{\dot\a},.\}$ be any $\hs$-valued frame.
Consider some point $\xi\in\cM^{3,1}$, and assume that  ${e_{\dot\a}^{\ \mu}}|_\xi$
is invertible.
Choose shifted coordinates $y^\mu = x^\mu - \xi^\mu$
which are centered at $\xi$, i.e.
\begin{align}
 y^\mu|_\xi = 0 \ .
\end{align}
Then we can construct local coordinates $\tilde y^\mu$ around $\xi$ of the form
\begin{align}
 \tilde y^\mu =  \phi^\mu_\s(u)\, y^\s + \phi^\mu_{\a{\dot\b}}(u) y^{\a} y^{{\dot\b}}
\label{tilde-x-lin}
 \end{align}
such that the frame ${\tilde e}_{\dot\a}^{\ \mu} = \{{\bf T}_{\dot\a},\tilde y^\mu\}$ satisfies
\begin{align}
{\tilde e}_{\dot\a}^{\ \mu}\big|_\xi &= \d^\mu_{\dot\a}  ,
\qquad  \g^{\mu\nu}|_\xi = \eta^{\mu\nu}\nn\\
\del_\s \g^{\mu\nu}\big|_\xi &= 0
 \label{frame-normal}
\end{align}
at  $\xi \in\cM^{3,1}$.
\label{lem:normal-coords-hs}

\vspace{0.2cm}

In other words,
all $\hs$ components of the frame, the metric and the Levi-Civita connection (but not the Weitzenb\"ock torsion) can be  absorbed locally by
this change of coordinates, corresponding to a local inertial system.
The proof is straightforward, closely following the usual steps in establishing local normal coordinates in GR. Alternatively, one can also achieve $\del_\s G^{\mu\nu}\big|_\xi = 0$. However, $\del {\tilde e}_{\dot\a}^{\ \mu}\big|_\xi = 0$ would be asking too much, because this determines the torsion tensor \eqref{Weitzenbock-torsion-def}.

At the linearized level, the elimination
of the $\hs$-valued frame components
${e}_{\dot\a}^{\ \mu} = {\bar e}_{\dot\a}^{\ \mu} 
+ \d {e}_{\dot\a}^{\ \mu}$ with $\d {e}_{\dot\a}^{\ \mu} \in \hs$
can be achieved with $\hs$-valued coordinates whose $\cC^0$ part is undeformed, i.e.
\begin{align}
 \tilde y^\mu := y^\mu + O(\d e y) \ .
 \label{tilde-y-local-lin}
 \end{align}
 Here $y^\mu = x^\mu -\xi^\mu$ are shifted global  Cartesian coordinates around the reference point $\xi$,
 and the $\hs$ components vanish at $\xi$.
  Indeed, for the new  coordinates $\tilde y^\mu = \phi^\mu_\nu y^\nu$
  with
  \begin{align}
 \phi^\mu_\s := {\bar e}_{\dot\a}^{\ \mu}{e}_{\s}^{\ \mu} \big|_\xi^{-1}
 = \d^\mu_\s + O(\d e) \
\end{align}
(which implies  \eqref{tilde-y-local-lin})  the frame reduces to 
\begin{align}
 {\tilde e}_{\dot\a}^{\ \mu}\big|_\xi
 &= \phi^\mu_\s\, {e}_{\a}^{\ \s}|_\xi
   = {\bar e}_{\dot\a}^{\ \mu} \ .
\end{align}
Hence the effective geometry near $\xi$ can be extracted simply by
projecting to the  $\cC^0$ component of the frame, in the linearized regime. This is no longer restricted to a particular point but extends to a coordinate patch small compared to the curvature scale.
The global geometry can then be obtained as usual by patching together local $\hs$-valued coordinates.

\subsection{Gauge transformations and volume-preserving $\hs$ diffeos}
\label{sec:gauge-trafos-diffeos}

On the 6-dimensional symplectic bundle $\cB$ over spacetime $\cM^{3,1}$, gauge transformations can be interpreted as  Hamiltonian vector fields $\{\Lambda,.\}$
\begin{align}
\label{gauge-trafo}
    \phi \to \phi + \{\Lambda,\phi\} \ .
\end{align}
This is a subset of volume-preserving vector fields on $\cB$, which provides enough degrees of freedom to describe all volume-preserving diffeomorphisms on 3+1-dimensional spacetime\footnote{as well as certain higher-spin generalizations thereof.} $\cM^{1,3}$.
Indeed for classical functions $\phi = \phi(y)$, the above transformation acts as
\begin{align}
\label{gauge-trafo-functions}
    \phi \to \phi + \{\Lambda,y^\mu\} \del_\mu \phi \ 
\end{align}
which is a ($\hs$-deformed) diffeomorphism along the vector field
\begin{align}
 V^\mu = \{\Lambda,y^\mu\} \ .
\end{align}
The resulting vector field $V^\mu$ on spacetime is volume-preserving, since
\begin{align}
\label{div-constraint}
 \del_\mu V^\mu =
    \del_\mu\{y^\mu,\Lambda\} \sim \del_\mu (\theta^{\mu\nu}\del_\nu \Lambda) 
    \approx \theta^{\mu\nu}\del_\mu\del_\nu \Lambda = 0 \ 
\end{align}
for any $\Lambda$, assuming local normal coordinates where $\theta^{\mu\nu}$ can be treated as constant; the precise statement is 
$\del_\mu(\rho_M V^{\mu}) = 0$
\cite{Steinacker:2024unq}.
Conversely, any volume-preserving vector field on spacetime can be reconstructed in this way, as 
discussed in Appendix \ref{sec:hs-Lorentz}.
These canonical transformations on the bundle act
as vector fields on spacetime functions 
and as Lie derivatives on spacetime tensors.
 For example, frame and metric transform as \cite{Steinacker:2020xph}
\begin{align}
 \delta_\Lambda E^{\a\mu} = \cL_V E^{\a\mu} , \qquad 
  \delta_\Lambda G^{\mu\nu} = \cL_V G^{\mu\nu} \ .
 \label{vectorfield-Lie}
\end{align}
However
$V^\mu$ and the above expressions   typically have some $\hs$ components.
We shall therefore denote them as {\bf $\hs$-deformed} vector fields or diffeos\footnote{It turns out that the $\hs$ components are uniquely specified by the classical components, hence the information is the same and the name ``deformed'' is appropriate. Moreover, the transformations act in a non-standard way on the $\hs$ generators.
This is somewhat analogous to classical mechanics, where general transformations $\d q^i(q)$ in configuration space are generated by Noether charges via $\{\d q^i(q) \ p_i,.\}$ acting on $q^i$, while their action on $p_j$ is more complicated.}.
In this sense, any volume-preserving vector field on $\cM^{3,1}$ can be obtained
from gauge transformations. These 
 are exact symmetries of the action even at the quantum level, and are expected to imply more-or-less the standard consequences {\em for the local, tensorial part of the effective action}:

\paragraph{Significance of $\hs$-deformed smmetries}

We have seen that
the generalized diffeomorphisms typically include some $\hs$ deformation.
Nevertheless, since the action involves a projection to the classical (non-$\hs$) part,
they imply that the effective (tensorial) action for $\cC^0$-valued fields  is invariant in the standard sense:
\begin{align}
    0= \int \delta_\Lambda \cL[\phi] = \int [\delta_\Lambda \cL[\phi]]_0 \ =  \int \frac{\del}{\del\phi}\cL[\phi][\delta_\Lambda\phi]_0 \ ,
\end{align}
where $\phi$ indicates classical tensorial  fields and $\cL$ denotes some Lagrangian density. Hence the action is invariant under the classical transformations $[\delta_\Lambda\phi]_0$, which includes 
local Lorentz transformations as a part of volume-preserving diffeos.
For the frame $e^{\dot\a\mu} = \{t^{\dot\a},y^\mu\}$, this can be compensated by the action of the global $SO(3,1)$ on the frame index $\dot\a$, so that the frame is invariant after projection to $\cC^0$.
More generally, all physical tensor fields  transform tensorially (i.e. via Lie derivatives $\cL_V$)
under volume-preserving diffeos
as long as their $\hs$ components are negligible, which is expected to hold in the locally flat regime\footnote{This is elaborated in \cite{Gass:2025bqr}, including estimates for the $\hs$ contributions. For a general discussion see \cite{Steinacker:2024unq}.}.

The effective metric $G^{\mu\nu}$ is the only classical background tensor in the semi-classical YM action, governing all fluctuation modes. 
In that sense, local Lorentz invariance {\em emerges} in the low-energy local physics\footnote{A situation where local Lorentz invariance is broken is found in the extreme IR regime of gravity, which is not described by a local tensorial action.}, as in GR. 
The background does admit the Lorentz-violating tensor given by the time-like FLRW vector field $\cT$; however, it does not seem to show up in the low-energy local action.


\subsection{$\hs$- deformed diffeos and Lorentz transformations}
\label{sec:hs-Lorentz}

\paragraph{Local reconstruction of  divergence-free vector fields.}

It is shown in \cite{Steinacker:2024unq} that 
all volume-preserving vector fields $V$ on $\cM^{1,3}$ can be generated as $V = [\{\Lambda,.\}]_0$  by some uniquly determined $\Lambda = A^\sigma(y) u_\sigma$. This is easy to see 
in the locally flat regime using local Cartesian coordinates $y^\mu$, where this ansatz generates the vector field 
\begin{align}
\label{local-frame-full}
 V^\mu &=  \{\Lambda,y^\mu\}
     = A^{\sigma} \{ u_\sigma,y^\mu\} +  \{A^{\sigma} ,y^\mu\} u_\sigma
       \ .
\end{align}
For vector fields $A^\sigma(y)$
with wavelengths $\ll L_{\rm IR}$, the second term dominates, and using
the bracket relations \eqref{y-u-brackets} we find
\begin{align}
\{\Lambda,y^i\}
      & \approx \{A^{l} ,y^i\} u_l
      = L_{\rm NC}^2\del_0 A^{l} u^i u_l   \nn\\
\{\Lambda,y^0\}
      &\approx \{A^{l} ,y^0\} u_l
      = - L_{\rm NC}^2 \del_j  A^{l} u^j u_l \ .
      \label{frame-full-local}
\end{align}
Projecting to the spin 0 sector using \eqref{u-average},
 one finds
\begin{align}
\label{reconstruction-equation-local}
\begin{aligned}
\ [V^{0}]_0
 &=  -\frac{1}{3}L_{\rm IR} \del_i A^{i} \\
\ [V^k]_0
 &\approx  \frac{1}{3} L_{\rm IR} \del_0 A^{k}
 \end{aligned}
\end{align}
dropping $A^0$ (in unitary gauge) and
replacing $L_{\rm NC}^2 = \xi^0 = L_{\rm IR}$
in units with $r=1$.
The resulting vector field $V^\mu$ is  automatically divergence-free, and for any divergence-free vector field $V$ one
can always find appropriate $A^i$,
which is unique up to a divergence-free static vector field.
However for very slowly varying vector fields $V$  such as generators of global symmetries,
the first term in \eqref{local-frame-full} cannot be neglected, and
the generator $\Lambda$ must be determined more carefully.

\paragraph{Example: local rotations and $\hs$-deformed Lorentz boosts.}

The $SO(3)$ rotations around the reference point $\xi$ are
examples of volume-preserving diffeos where the first term in \eqref{local-frame-full} cannot be neglected. They are given by\footnote{This holds in the minimal case $n=0$ \cite{Manta:2025inq}, and generically at late times.} 
\begin{align}
  \cM^{(ij)} =  r^{-1}(u^i y^j - u^j y^i) \ .
\end{align}
To verify this, consider 
\begin{align}
 r \{\cM^{(ij)},y^k\} &= \{u^i y^j - u^j y^i,y^k\} \nn\\
  &= r u^i(y^j u^k - y^k u^j) 
  + r y^j(\d^{ik} - u^i u^k) -  (i\leftrightarrow j)   \nn\\
   &= r (\d^{ik} y^j - \d^{jk} y^i) 
\end{align}
and 
\begin{align}
 r \{\cM^{(ij)},u^k\} &= \{u^i y^j - u^j y^i,u^k\} \nn\\
  &= u^i\{ y^j,u^k\} +  y^j\{u^i,u^k\} -  (i\leftrightarrow j)   \nn\\
   &= r (\d^{ik} u^j - \d^{jk} u^i) \ .
\end{align}
One can check that $\{\cM^{(ij)},y^0\} = 0 = \{\cM^{(ij)},u^0\}$.
The $so(3)$ Lie algebra for the Poisson brackets $\{\cM^{(ij)},\cM^{kl}\}$ follows immediately.

\paragraph{Lorentz boosts:}

The local Lorentz  boosts turn out to be generated by
\begin{align}
 \label{Loentz-boosts}
 \cM^{(0i)} = r^{-1}\Big(\big(-\frac 3{2} y^0 + \frac{3}{4 }L_{\rm IR}\big)u^i + u^0 y^i\Big) \ .
\end{align}
One can verify with some effort that they indeed generate  the standard Lorentz transformations upon projecting to
$\cC^0$ in the locally flat regime:
\begin{align}
\label{action-Mmunu-class}
    [\{\cM^{(\mu\nu)}, y^\rho\}]_0 \sim \eta^{\mu\rho} y^\nu - \eta^{\nu\rho} y^\mu \ .
\end{align}
This requires dropping explicit $u^0$ on the rhs, which locally vanish as functions on the bundle space.
On the other hand, they  act in a non-covariant way on the $u^\mu$ generators:
\begin{align}
 \{\cM^{(0i)},u^j\}
 &\sim 0   \nn\\
  \{\cM^{(0i)},u^0\} &\sim -\frac 3{2} u^i
\end{align}
dropping again $u^0$. 
The Poisson brackets between the rotation generators
satisfy the usual $so(3)$ Lie algebra.
By construction, this also holds for $\{\cM^{(ij)},\cM^{(0k)}\}$.
In contrast, the brackets between  boost generators satisfy
\begin{align}
\label{brackets-boosts}
     \{\cM^{(0i)},\cM^{(0j)}\}
       = -\frac 32 \cM^{(ij)} 
\end{align}
dropping $u^0$.
Nevertheless, the Lorentz-transformations of physical spacetime tensor fields (such as the graviton) is the standard one as discussed above, as long as they are classical.


\subsection{Dilaton contributions}
\label{sec:appendix-dilaton}

We need
\begin{align}
\label{del-rho2}
\rho^2 &= \rho_M \det e^{\dot\a\mu} 
 = \bar\rho^2 \det(\delta^\mu_\nu +  \bar e_{\dot\a\nu} \d e^{\dot\a\mu} ) 
\end{align}
in Cartesian coordinates. We use $\bar e^{\dot\a\nu} = \a\sinh(\tau)\eta^{\dot\a\nu}$ and the expansions
\begin{align}
  \det(1+B)^{-1} &= 1- \tr(B) + \frac 12\big( (\tr B)^2 + \tr (B^2)\big) + O(B^3)   
\end{align}
for $B^\mu_\nu = \bar e_{\dot\a\nu} \d e^{\dot\a\mu}$, where 
 $\d e^{\dot\a\nu} =  r A^{\dot\a \mu} \big(u_\mu k_\sigma (\xi^\sigma u^\nu -  u^\sigma \xi^\nu) 
       + (\delta^\nu_\mu - u^\nu u_\mu)\big)$
in the plane wave expansion \eqref{frame-modes} for the matrix fluctuations $\cA$.
Recalling that time-like components for physical modes vanish  $\xi_{\dot\a} A^{\dot\a\mu} = 0$ \eqref{spacelike-A}, we find
\begin{align}
  \tr(B) &=  \tr(\bar e_{\dot\a\nu} \d e^{\dot\a\mu})
   = \frac{r }{\a \sinh(\tau)}  ( (k_\sigma \xi^\sigma -1) A^{\dot\a \mu} u_{\dot\a} u_\mu  + A)\nn\\
  [\tr(B)]_0  &= \frac{r}{\a\sinh(\tau)} \big(\frac 13A^{\dot k j} \d_{kj} (k_\sigma \xi^\sigma-1) + A^{\dot \a\mu} \eta_{\dot\a\mu}\big) 
  \ \sim \ \frac {r^2 k_0}{3\a} \tr(A)
\end{align}
(the last form holds for $k_0 L_{\rm IR} \gg 1$).
We use the notation 
\begin{align}
 \tr(A) &= A^{\dot\a \mu} \eta_{\dot\a \mu} =  A^{\dot k j} \delta_{\dot k j}, 
    \qquad k\cdot\xi = k_\sigma \xi^\sigma  \nn\\
 \tr(A A^T) &= A^{\dot k i} A^{\dot l j} \d_{\dot k\dot l} \d_{ij} , \qquad
 \tr(A A) = A^{\dot k i} A^{\dot l j} \d_{\dot k j} \d_{\dot l i}
\end{align}
at or near the reference point $\xi$.
Note that this depends 
only on the frequency $k_0$, but not on the space-like momentum.
Similarly,
\begin{align}
(\tr(B))^2 &=  \frac{r^2}{\a^2\sinh^2(\tau)} 
\big((k\cdot \xi -1) A^{\dot\a \mu}  u_{\dot\a} u_\mu  + \tr(A)\big)
\big((k\cdot \xi -1)A^{\dot\b \nu}u_{\dot\b} u_\nu  + \tr(A)\big)
\nn\\
%
[(\tr(B))^2]_0 
 &= \frac{r^2}{\a^2\sinh^2(\tau)} 
\Big(\frac 23\big(k\cdot \xi + \frac 12\big) (\tr(A))^2  + \frac{1}{15}(k\cdot \xi -1)^2 \big(\tr(A A^T) 
     + \tr(A A) + (\tr(A))^2\big)\Big) \nn\\
  \tr(B^2) &= \tr(\bar e_{\dot\a\nu} \d e^{\dot\a\mu} \bar e_{\dot\b\mu} \d e^{\dot\b\nu} )  \nn\\
%
 &= \frac {r^2}{\a^2\sinh^2(\tau)}
      A^{\dot\a \r}\Big( (k\cdot \xi-1) u_\r u_{\dot\b}
        + \eta_{\dot\b\rho}\Big)
      A^{\dot\b \t}\Big( (k\cdot \xi -1) u_\t u_{\dot\a} + \eta_{\dot\a\t}\Big) \nn\\
[\tr(B^2)]_0 &=  \frac {r^2}{\a^2\sinh^2(\tau)}
      A^{\dot\a \r}A^{\dot\b \t}
      \Big(\frac 13\big(2 k\cdot \xi+1\big)\delta_{\rho\dot\b}\delta_{\dot\a\t}
 + (k\cdot \xi-1)^2[u_\r u_{\dot\b}u_\t u_{\dot\a}]_0
      \Big)  \nn\\
&=  \frac {r^2}{\a^2\sinh^2(\tau)}
      \Big(\frac 23\big(k\cdot \xi+\frac 12\big) \tr(AA)
 + \frac 1{15} (k\cdot \xi-1)^2
     \big(\tr(AA) + \tr(AA^T) + (\tr(A))^2 \big)
      \Big) 
\end{align}
 leads to the following contributions of the physical fluctuation modes to vacuum energy:
\begin{align}
\bar\rho^{2}[\rho^{-2}]_0 
&\approx 1 - \frac{r^2 k_0}{3\a}  \tr(A)
+ \frac {(r^2 k_0)^2}{15\a^2}
     \big(\tr(AA) + \tr(AA^T) + (\tr(A))^2 \big) \ + O(A^3) \ .
    \label{rho-2-expand-A}
\end{align}
This provides the term $V_{01} k_0^2 \tr(A A^T)$ with $V_{01} = \frac{r^4}{15\a^2}>0$ in \eqref{S-grav-quad-full} needed to stabilize the would-be tachyonic modes; note that the mass terms are suppressed at late times.
Clearly the $\hs$ components are essential here.


\subsection{Non-derivative contributions from the YM term}
\label{sec:appendix-YM-mass}

Keeping the non-derivative terms, the Yang-Mills term \eqref{YM-action-quad} acquires an extra contribution:
\begin{align}
\label{F-A-explicit-fill}
  \cF^{\dot\a\dot\b} &=  \a (- \{t^{\dot\a},A^{\dot\b \mu}u_\mu\}
   + \{t^{\dot\b},A^{\dot\a \mu}u_\mu\})  \ 
    -\{A^{\dot\a\nu} u_\nu,A^{\dot\b \mu}u_\mu\} \nn\\
    &\approx  \frac{\a}{r^3\sinh(\t)} (- A^{\dot\b \mu}\theta^{\dot\a\nu}
    + A^{\dot\a \mu}\theta^{\dot\b\nu})\eta_{\mu\nu}
    + O(\del A) + O(A^2)
 \end{align}
 using 
 \begin{align}
      \{t^{\dot\a},u^\mu\} 
       &\approx -\frac{1}{r^3\sinh(\t)}\theta^{\dot\a\mu}
      = \frac{1}{r^2\sinh(\t)}(u^{\dot\a} \xi^\mu - u^\mu \xi^{\dot\a}) \ .
 \end{align}
 This leads to extra terms
 \begin{align}
     \cF^{\dot\a\dot\b}\cF_{\dot\a\dot\b} \
      &\supset \frac{2\a^2}{r^4\sinh^2(\t)}
     \Big( A^{\dot\b \mu}u^\nu \xi^{\dot\a}\eta_{\mu\nu} A^{\dot\b \rho}u^\sigma \xi^{\dot\a}\eta_{\rho\sigma}\Big) 
      = -\frac{2\a^2}{r^2}
      A^{\dot\b \mu}u_\mu  A^{\dot\b \rho}u_\rho \nn\\
[\cF^{\dot\a\dot\b}\cF_{\dot\a\dot\b}]_0 \
     &\supset \ -\frac{2\a^2}{3r^2}
      A^{\dot\b i}  A^{\dot\b j}\delta_{ij}
      \label{mass-term-YM}
 \end{align}
 recalling $\xi_{\dot\a} A^{\dot\a\mu} = 0 = A^{\dot\a\mu}\xi_\mu$ for the physical modes, and $\xi_\mu \xi^\mu = - r^2\sinh^2(\tau)$. 
 Alternatively this can be obtained using
 $\Box u^\mu = -\frac {\a^2}{3r^2} u^\mu$ (see (B.5) in \cite{Steinacker:2023cuf}).
 This leads to the mass term $V_{00}\tr(A A^T)$ with $V_{00} = -\frac {\a^2}{r^2} < 0$ in \eqref{S-grav-quad-full} needed to stabilize the would-be tachyonic modes.


\subsection{Cosmic scale function}
\label{sec:cosm-scale}

Consider the effective metric \eqref{eff-metric-App}
\begin{align}
G_{\mu\nu} &=  \frac{\a^{2}}{r^4} e^{\t} \eta_{\mu\nu} =: \frac 1{A(\t)} \eta_{\mu\nu} , \qquad A = r^{4}\a^{-2} e^{-\t}
\end{align}
in Cartesian coordinates. We can cast this into $k=-1$ FLRW form  as follows
\begin{align}
 d s^2_G = G_{\mu\nu} d x^\mu d x^\nu 
&= - r^2 A^{-1} \sinh^2\tau \, d \tau^2 + r^2 A^{-1} \cosh^2\tau\, d \Sigma^2,
\label{eff-metric-FRW}
\end{align}
(observe that this depends only on $\frac{\a}{r}$, as it should), where 
\begin{align}
    d\Sigma^2 = d\chi^2 + \sinh^2\chi (d\theta^2 + \sin^2 \theta \, d\varphi^2)
    \label{dSigma2}
\end{align}
is the invariant length element on the spacelike hyperboloids $H^3$. 
 We can bring it to the standard FLRW form 
\begin{align}
 d s^2_G = -d t^2 + a(t)^2 d\Sigma^2
 \label{FRW-standard-metric}
\end{align}
via 
\begin{align}
dt^2 &= r^2 A^{-1} \sinh^2\tau \,   d \tau^2,
\label{dt-d-eta}
\\
a(t)^2 &= r^2 A^{-1} \cosh^2\tau \ \sim \frac{\a^2}{r^2} e^{3\tau} \ .
\label{relation-a-t-eta}
\end{align}
The Hubble scale is found to be
\begin{align}
\label{Hubble-scale-1}
   H(t) &= \frac{\dot a}{a} 
    = r\a^{-2} e^{-3/2\t} 
    (\frac{d\a}{d\t} + \frac 32 \a) 
\end{align}
which for $\a = e^{\varepsilon\tau}$ gives 
\begin{align}
\label{Hubble-scale-app}
    H(t) &= r(\varepsilon + \frac 32) e^{-(3/2 + \varepsilon)\t} 
    = (\varepsilon + \frac 32) \frac 1{a(t)} \propto  \frac 1t \ .
\end{align}
A particularly interesting case is $\a \sim e^{-3/4\t}$, where $\rho = const$ and $a(t) \sim e^{3/4\t}$ \cite{Manta:2025tcl}.

\subsection{Modified d'Alembertian $\tilde\Box$.}

Recalling $\g^{\mu\nu} = \a^{2} e^{2\tau} \eta^{\mu\nu}$, we compute the modified d'Alembertian $\tilde\Box$ \eqref{Box-tilde-def}
\begin{align}
\label{tilde-Box-planewave}
    \tilde \Box e^{ikx}
    &:=  r^{-4} \a^2 \{x^\mu,\gamma_{\mu\nu}\{x^\nu,e^{ikx}\}\} \nn\\
    &= r^{-4}e^{-2\tau} \{x^\mu,\eta_{\mu\nu}\{x^\nu,e^{ikx}\}\}
     \approx  - r^{-4}e^{-2\tau}\eta_{\mu\mu'}\theta^{\mu\nu} \theta^{\mu'\sigma} k_\sigma k_\nu e^{ikx} \nn\\
    &= - r^{2} \frac{e^{-2\tau}}{\xi_4^2}(x^\mu t^\nu - x^\nu t^\mu)(x_\mu t^\sigma - x^\sigma t_\mu)
     k_\sigma k_\nu e^{ikx} \nn\\
    &\approx - r^{2} e^{-2\tau} (- t^\nu t^\sigma + r^{-4} x^\nu  x^\sigma )
     k_\sigma k_\nu e^{ikx} \nn\\
     &= (u^\nu u^\sigma k_\sigma k_\nu - k_0^2)  e^{ikx} \ 
\end{align}
in Cartesian coordinates (or in analogous local normal coordinates).
This decomposes into a classical part
\begin{align}
\label{tilde-Box-planewave-2}
  [\tilde \Box e^{ikx}]_0 &= (\frac {1}{3} \vec k^2 
    - k_0^2) e^{ikx}
\end{align}
and a $\hs$-valued part, which is typically of the same size as the classical part.
Comparing this with the metric d'Alembertian of the cosmic background
\begin{align}
    \Box_G \sim G^{\mu\nu} \del_\mu \del_\nu
     =  r^{4}\a^{-2} e^{-\t} \eta^{\mu\nu}\del_\mu \del_\nu
\end{align}
we see that the classical sector of $\tilde\Box$ reduces approximately to the metric d'Alembertian with speed of light reduced by a factor $\frac 1{\sqrt{3}}$:
\begin{align}
\label{Box-tilde-Box}
    \tilde\Box \sim  \frac 13r^{-4}\a^{2} e^{\t}\Box_{G,c\to\frac 1{\sqrt{3}} c} \ .
\end{align}

\subsection{Harmonic gauge condition}
\label{sec:harmonic-gauge}

We derive the harmonic gauge condition for the trace-reversed metric fluctuation,  provided the torsion of the background is negligible.
Observe first that 
the trace-reversed metric perturbation can be written as
\begin{align}
    \delta \bar G^{\mu\nu} 
    &\equiv \d G^{\mu\nu} - \frac 12
     G^{\mu\nu}(G_{..}\d G^{..}) \nn\\
    &= \rho^{-2} \d e^{\dot\a\mu} e^{\dot\a\nu}
    + \rho^{-2} e^{\dot\a\mu} \d e^{\dot\a\nu}
    - G^{\mu\nu}\big(\rho^{-2}\delta\rho^2
    + \frac 12
     G_{..}\d G^{..}\big) \nn\\
    &= \rho^{-2} \d e^{\dot\a\mu} e^{\dot\a\nu}
    + \rho^{-2} e^{\dot\a\mu} \d e^{\dot\a\nu}
\end{align}
using 
$\rho^{-2}\delta\rho^2 = -\frac 12 G_{..}\delta G^{..}$
which follows from \eqref{dilaton-def}.
Now we assume that the Weitzenböck torsion $T^{\dot\a\dot\b\mu}$ \eqref{Weitzenbock-torsion-def} of the background is sufficiently small, so that $\del_\nu e^{\dot\a\mu}$ and $\del\rho$ can be neglected in Riemann normal coordinates $\nabla^{(G)}_\mu = \del_\mu$. Then we can proceed  as 
\begin{align}
    \nabla^{(G)}_\mu \delta \bar G^{\mu\nu}  &= \nabla^{(G)}_\mu \big(  \d G^{\mu\nu} - \frac 12
     G^{\mu\nu}(G_{..}\d G^{..}) \big) \nn\\
    &= \nabla^{(G)}_\mu\big(  \rho^{-2} \d e^{\dot\a\mu} e^{\dot\a\nu}
    + \rho^{-2} e^{\dot\a\mu} \d e^{\dot\a\nu}  \big)\nn\\
    &\approx \rho^{-2} e^{\dot\a\mu}  \partial_\mu \d e^{\dot\a\nu} \approx 0
\end{align}
using the divergence constraint \cite{Fredenhagen:2021bnw}
\begin{align}
\del_\mu(\rho_M e^{\dot\a\mu}) = 0
\label{div-free}
\end{align}
which is a consequence of the Jacobi identity, and
\begin{align}
    e^{\dot\a\mu} \del_\mu\d e^{\dot\a\nu}
 &= \{\TT^{\dot\a},\{\d \TT^{\dot\a},x^\nu\}\} \nn\\
 &= -\{\d \TT^{\dot\a},\{x^\nu, \TT^{\dot\a}\}\}
  - \{x^\nu,\{\TT^{\dot\a},\delta\TT^{\dot\a}\}\} \nn\\
   &= \{\d \TT^{\dot\a}, e^{\dot\a\mu}\}
    \ \propto L_{\rm NC}^2\del\delta \TT^{\dot\a}\, T^{\dot\a . .} \approx 0 \ .
\end{align}
Here we use the Lorentz gauge condition $\{\TT^{\dot\a},\delta\TT^{\dot\a}\} = 0$ for the matrix fluctuations, which is used in the quantization in complete analogy to Yang-Mills theory, cf \cite{Steinacker:2023myp}.  $L_{\rm NC}$ denotes the length scale of noncommutativity.  Hence if the background has negligible torsion, then the harmonic gauge condition holds to a good approximation.



\end{document}